\documentclass{ws-rv9x6}
\usepackage{subfigure}   
\usepackage{ws-rv-thm}   
\usepackage{ws-rv-van}   

\usepackage[utf8]{inputenc}
\usepackage[greek,ukrainian,english]{babel}
\usepackage{wrapfig}
\usepackage{wasysym}
\usepackage{float}

\usepackage{lmodern}
\def\eps{\varepsilon}
\def\ukrtext#1{\selectlanguage{ukrainian}#1\selectlanguage{english}}

\newsavebox{\foobox}
\newcommand{\slantbox}[2][.5]
  {%
    \mbox
      {%
        \sbox{\foobox}{#2}%
        \hskip\wd\foobox
        \pdfsave
        \pdfsetmatrix{1 0 #1 1}%
        \llap{\usebox{\foobox}}%
        \pdfrestore
      }%
  }

\begin{document}

\chapter[Approaches to the Classification of Complex Systems: Words, Texts, and More]{Approaches to the Classification of Complex Systems: Words, Texts, and More\label{ra_ch1}}

\author[A. Rovenchak]{Andrij Rovenchak\footnote{ORCID: \url{https://orcid.org/0000-0002-0452-6873}}}

\address{Ivan Franko National University of Lviv,\\
andrij.rovenchak@lnu.edu.ua, \ andrij.rovenchak@gmail.com
}

\begin{abstract}
The Chapter starts with introductory information about quantitative linguistics notions, like rank--frequency dependence, Zipf's law, frequency spectra, etc. Similarities in distributions of words in texts with level occupation in quantum ensembles hint at a superficial analogy with statistical physics. This enables one to define various parameters for texts based on this physical analogy, including ``temperature'', ``chemical potential'', entropy, and some others. Such parameters provide a set of variables to classify texts serving as an example of complex systems. Moreover, texts are perhaps the easiest complex systems to collect and analyze. 

Similar approaches can be developed to study, for instance, genomes due to well-known linguistic analogies. We consider a couple of approaches to define nucleotide
sequences in mitochondrial DNAs and viral RNAs and demonstrate their possible
application as an auxiliary tool for comparative analysis of genomes.

Finally, we discuss entropy as one of the parameters, which can be easily computed
from rank--frequency dependences. Being a discriminating parameter in some problems
of classification of complex systems, entropy can be given a proper interpretation only in
a limited class of problems. Its overall role and significance remain an open issue so far.\\
\end{abstract}


\body

\tableofcontents

\bigskip
\section{Introduction}

\begin{wrapfigure}{r}{0.335\textwidth}
\vspace*{-3ex}
\includegraphics[clip,height=0.216\textheight=1.0]{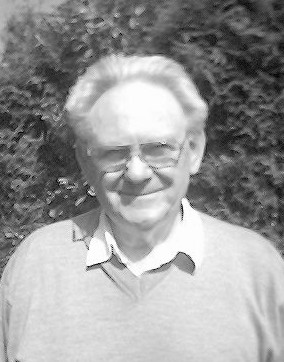}
\begin{centering}
{\footnotesize\qquad Gabriel Altmann\\[-0.8ex]
\hspace*{0.6em}
(24.05.1931 -- 02.03.2020)}

\vspace*{1pt}
\hspace*{0.4em}{\scriptsize Source: Altmann's homepage

}
\vspace*{-3ex}
\end{centering}
\end{wrapfigure}
Approaches originating in natural sciences, physics in particular, have proven useful in other domains, including social sciences and humanities. Active development started in the 1990s\cite{note1}, however, sporadic earlier works can be found as well \cite{Simon:1957,Golomb:1962,Suppes:1970,Ijiri&Simon:1975,West&Salk:1987}.

The major part of this Chapter will focus on studies of texts as examples of complex systems. Before proceeding, it is thus worth mentioning a kind of warning once issued by Gabriel Altmann, a leading expert in quantitative linguistics \cite{Altmann:2008}:

\begin{quote}
As a matter of fact, the greatest problem is the undifferentiated identification of physical and linguistic entities\ldots\ 
One should not try to transfer the gravitation theory to dialectology (once made by a linguist!) or the complete thermodynamics to a linguistic discipline only because
the concept of entropy is applicable both in physics and linguistics. 
Thus, if there is some analogy between physical and linguistic entities\ldots\ 
there must exist a way to find it without reductionist attempts and respecting the thinghood of physical and linguistic entities. The only possibility is to take into account the properties of models, i.e. to scrutinize the common abstract super-systems -- if there are any.
\end{quote}

In what follows in this Chapter, I will try adhering to this caveat as much as possible. Sometimes, however, it would be too tempting to avoid such a step while analyzing complex systems \cite{Holovatch&Palchykov:2007,Vasilev_etal:2013,Rodriguez_etal:2014,Eroglu:2014}. Mathematical models and certain notions will be essentially borrowed from statistical physics and applied to objects and notions from beyond physics providing thus potentially new insights in these fields.
This is possible due to the fact that non-random systems of many entities are characterized by certain distributions. A general feature of this study is that some quantities being analogs of, e.\,g., temperature or entropy that are standard concepts in thermodynamics, can linked to those distributions in a way analogous to approaches of statistical mechanics.

The organization of the reminder of this Chapter is as follows. Upon finishing this Section with explanation of certain notions used throughout the text, I proceed with models used to describe rank--frequency dependences in \sref{sec:rf_models}. The notion of ``temperature'' in linguistics is discussed in \sref{sec:Temp} followed by argumentation toward its scaling in \sref{sec:Tscaling}. Linguistics analogies in genomic studies are described in \sref{sec:Genomes}. Results related to ``temperature'' are presented in \sref{sec:T-res} for two different problems. Section~\ref{sec:Entropy+params} addresses the notion of entropy and some other parameters with a potential use in the classification of complex systems, as demonstrated in \sref{sec:S-res} for two separate problems. Brief discussion in \sref{sec:Final} concludes the Chapter.

\bigskip
Let us start with some central notions used further. Consider the beginning of the famous Prince Hamlet's soliloquy,
\begin{quote}
  \begin{tabbing}
  To\quad\= be,\quad\= or\quad\= not\quad\= to\quad\= be,\quad\= 
  that\quad\= \,is\quad\= the\quad\= question\\
  (1)\>(2)\>(3)\>(4)\>(5)\>(6)\> \ (7)\>(8)\> \ (9)\> \ \ (10)
  \end{tabbing}
\end{quote}
Here, we have ten \textit{tokens}\index{token} (words) and eight \textit{types}\index{type} (different words) since tokens (1) and (5) as well as (2) and (6) are identical. Note that one can also apply the so called \textit{lemmatization}\index{lemmatization}, i.\,e., reducing words to their basic (dictionary) form called \textit{lemma}\index{lemma}. This lowers the number of types: token (8) `is' lemmatizes as `[to] be'. With this example we also see that, depending on the approach used, lemmatization can also lower the number of tokens: one can treat (1)--(2) as two tokens or as a single `to\_be'. Obviously, the things become even more complicated if several languages are considered. To avoid such unnecessary discussions, only non-lemmatized texts are analyzed hereafter. To be specific, the so called \textit{orthographic words}\index{orthographic word} are considered in this Chapter: texts are stripped of punctuation marks and alphanumeric sequences between two spaces are counted (the beginning and end of the line are treated as spaces). In case of bicameral scripts (those distinguishing upper- and lowercase letters, i.\,e., Roman, Cyrillic, Greek, and a few others), no case distinction is made. The total number of tokens will be denoted $N$ while the total number of types will be $V$.

Frequency lists are compiled in the form of pairs 
\[
<\!\textrm{token}\!>_i \quad f_i,
\]
where $f_i$ is the number of occurrences (absolute frequency, hence the notation) of $<\!\!\textrm{token}\!\!>_i$ in the given sample. Upon a descending sort of the list by the values of numbers $\{f_i\}$ (largest to smallest) we obtain the \index{rank--frequency dependence}\textit{rank--frequency dependence} $f_r$: the highest number corresponds to rank $r=1$, the second highest one has rank $r=2$ and so on. Tokens with equal frequencies occupy just a consecutive range of the respective ranks, cf. \cite[p.~13]{Baayen:2001}, \cite[p.~806]{Ludeling&Kyto:2009} or \cite[p.~10]{Popescu_etal:2009}. Such an approach is known as \index{ranking!ordinal}\textit{ordinal ranking}, as opposed to the so called \index{ranking!fractional}\textit{fractional} or \index{ranking!tied}\textit{tied ranking}, when a single value of the rank is assigned to such items, usually the mean of the rank range, though this is not a single possible option \cite{Kendall:1945}.

Inverting---in a certain sense---a rank--frequency dependence, one gets a \index{frequency spectrum}\textit{frequency spectrum}. The resulting set of quantities $N_j$ corresponds to the number or types having absolute frequency exactly equal to $j$.

\section{Rank--frequency dependence models}\label{sec:rf_models}

\begin{figure}
\centerline{\includegraphics[scale=0.872]{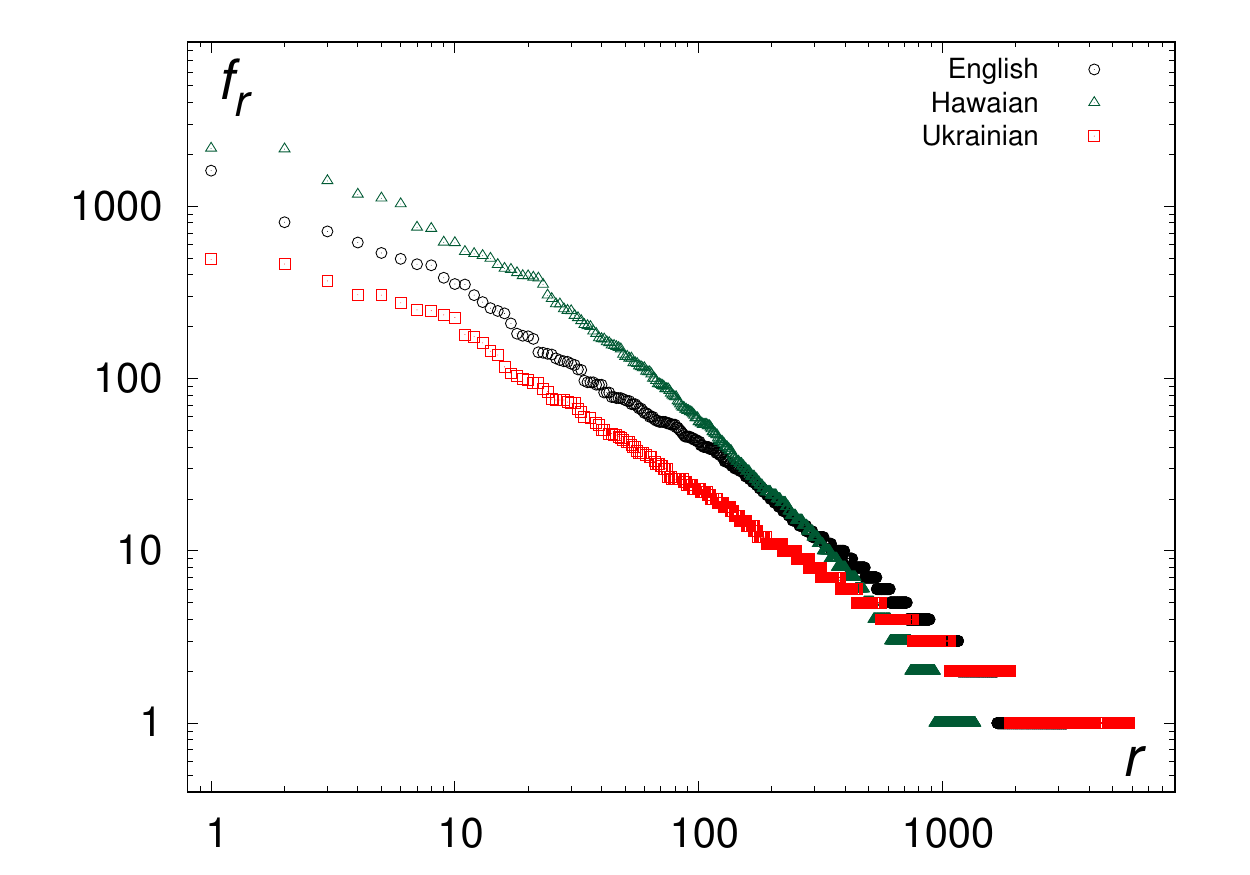}}
\vspace*{-1ex}
\caption{Rank--frequency dependences for \textit{Alice's Adventures in Wonderland} in English, Hawaian, and Ukrainian (own results). The relevant studies based in particular on this text were reported in Ref.~\refcite{Rovenchak:2015SLT,Rovenchak:2015CSCS}}
\label{fig:AW-RF}
\end{figure}

Figure~\ref{fig:AW-RF} demonstrates some typical rank--frequency dependences compiled from texts of Lewis Carroll's \textit{Alice's Adventures in Wonderland} in three languages.
Long plateaus at the lowermost absolute frequencies $f_r=1,2$, etc.\ mean a large number of the respective tokens. They are often referred to by special terms: tokens occurring only once in a sample are known as \index{hapax legomena}\textit{hapax legomena} (often just shortened to \index{hapaxes|see{hapax legomena}}\textit{hapaxes}) and those occurring twice are \index{dis legomena}\textit{dis legomena}. These are plurals of the respective Classical Greek terms {\greektext \textit{<'apax}} / {\greektext \textit{d'ic leg'omenon}} `[something] said [only] once / twice'. Subsequent terminology is straightforward but it will not be used here.

It is likely\cite{Petruszewycz:1973} that an (approximate) inverse proportionality between word's frequency and its rank was first observed in 1916 for French by Jean-Baptiste Estoup, a French stenographer \cite{Estoup:1916}. Similar results for English are known from no later than 1928, when the American nuclear physicist Edward Uhler Condon reported his findings \cite{Condon:1928}. These results are much better known from the works by George Kingsley Zipf, an American linguist \cite{Zipf:1935,Zipf:1949}, and the respective dependence is traditionally referred to as \index{Zipf's law}\textit{Zipf's law}:
\begin{align}\label{eq:Zipf}
f_r = \frac{A}{r^z}.
\end{align}
The exponent $z$ has values close to unity and depends in particular on language and genre \cite{Buk&Rovenchak:2004,Jayaram&Vidya:2008,Koplenig:2018,Wei&Liu:2019}. Originally, Zipf used $z=1$ for English.

Zipfian dependences are typical of complex systems of various sorts \cite{Gerlach_etal:2016,Holovatch_etal:2017}, not being unique to linguistic units \cite{Ferrer_i_Cancho&Sole:2001,Ha_etal:2002,Piantadosi:2014,Williams_etal:2015,Rovenchak&Buk:2018}. For instance, Zipf's law holds for city sizes based on census data \cite{Arcaute_etal:2014}, indirect estimation from number of customers in restaurants \cite{Ghosh_etal:2014}, and sizes of natural cities defined from nighttime satellite images \cite{Jiang_etal:2015}.

Zipf's law was also found in distributions of musical units \cite{Zanette:2006} and various units in genomic studies, including $k$-mers in genomes \cite{Sheinman_etal:2016} and abundances of expressed genes \cite{Furusawa&Kaneko:2003}. In economic context, it describes in particular wealth distribution \cite{Benisty:2017} and distribution of firm sizes in the United States \cite{Aoki&Nirei:2017}. 

The origins of power laws, of which Zipf's law is a particular example, has been examined from various viewpoints\cite{Mitzenmacher:2004,Newman:2005,Simkin&Roychowdhury:2011,Vasilev&Vasileva:2020} and it has been suggested that Zipf's law has a fundamental role in natural languages not being reproducible in random, so-called \index{monkey-typed text}\textit{monkey-typed}, texts\cite{Ferrer_i_Cancho&Sole:2002,Ferrer-i-Cancho&Elvevag:2010}.

Obviously, the rank--frequency distributions in reality are more complex than predicted by the simple Zipfian model \cite{Piantadosi:2014}. From Fig.~\ref{fig:AW-RF} we can observe that low ranks can significantly deviate from expected linear dependence when plotted in the log-log scale. To account for such situations, Beno\^\i{}t Mandelbrot \cite{Mandelbrot:1951} suggested the following generalization of (\ref{eq:Zipf})
\begin{align}\label{eq:ZM}
f_r = \frac{C}{(M+r)^B}
\end{align}
commonly known as the \index{Zipf--Mandelbrot law}\textit{Zipf--Mandelbrot law}. Similarly to Zipf's law, it is applicable in a vast variety of problems, reaching as far as, e.\,g., regularities observed in football matches\cite{Ramos_etal:2020}.

\bigskip
\noindent
\begin{minipage}{0.46\textwidth}
\centerline{\includegraphics[clip,height=0.22\textheight=1.0]{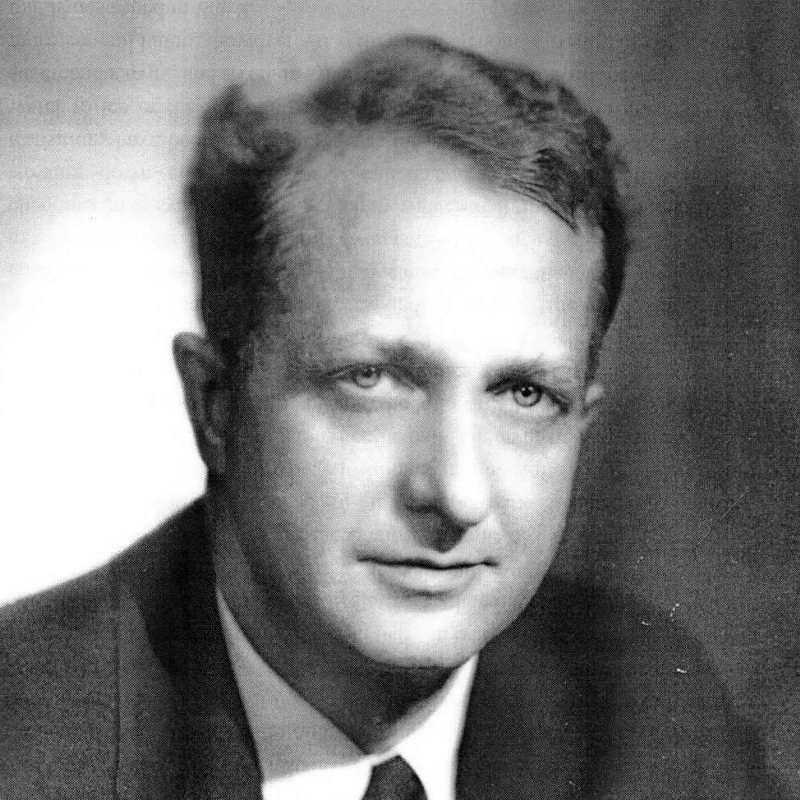}}
\centerline{{\footnotesize George Kingsley Zipf}}

\vspace*{-0.5ex}
\centerline{{\footnotesize (07.01.1902 -- 25.09.1950)}}

\smallskip
{\scriptsize Source:\quad https://peoplepill.com/people/\\ george-kingsley-zipf

}
\end{minipage}
\hfill
\begin{minipage}{0.46\textwidth}
\centerline{\includegraphics[clip,height=0.22\textheight=1.0]{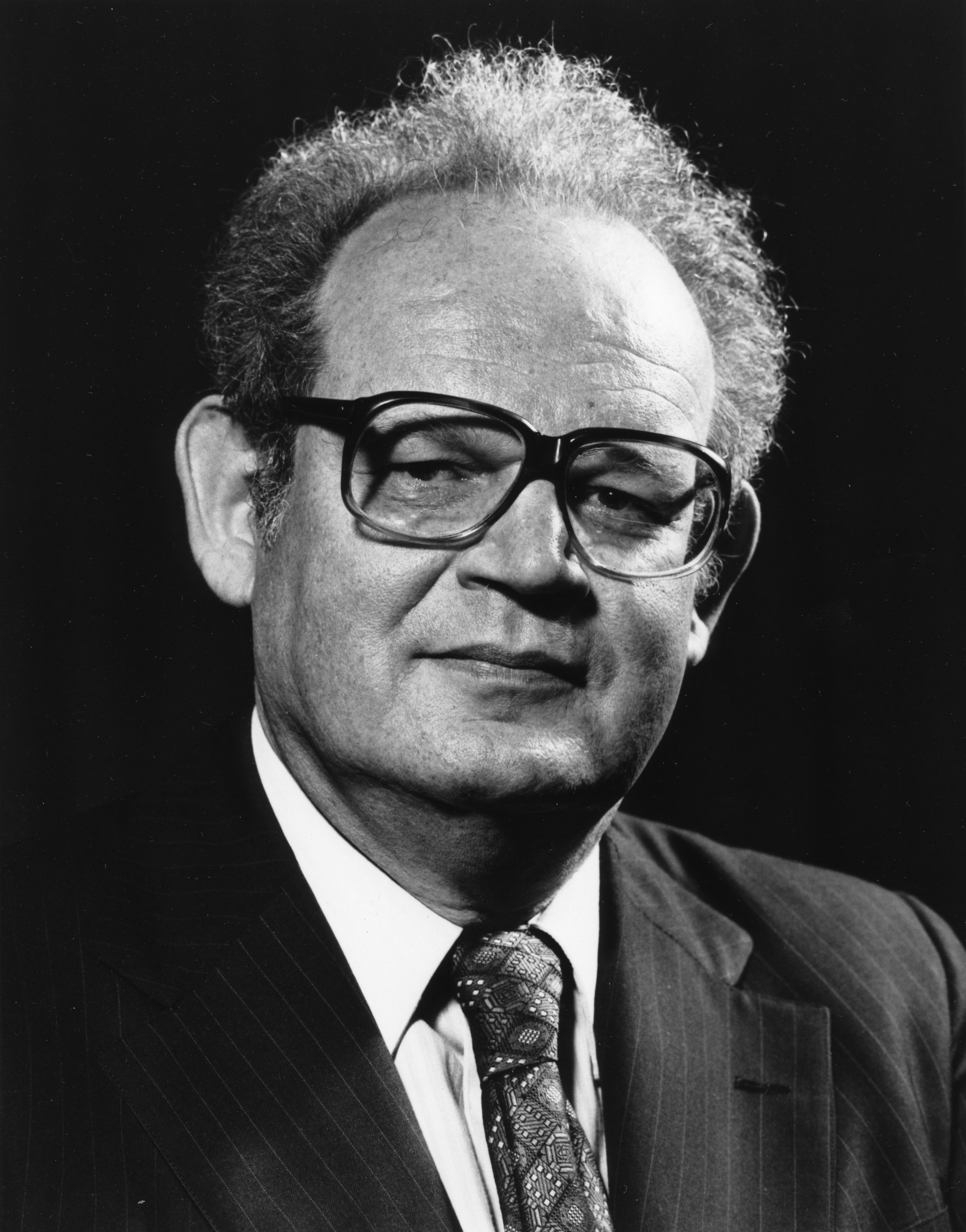}}
\centerline{{\footnotesize Beno\^it B. Mandelbrot}}

\vspace*{-0.5ex}
\centerline{{\footnotesize (20.11.1924 -- 14.10.2010)}}

\smallskip
{\scriptsize Courtesy: AIP Emilio Segr\`e Visual Archives, Physics Today Collection

}
\end{minipage}

\bigskip

\begin{figure}
\centerline{\includegraphics[scale=0.8]{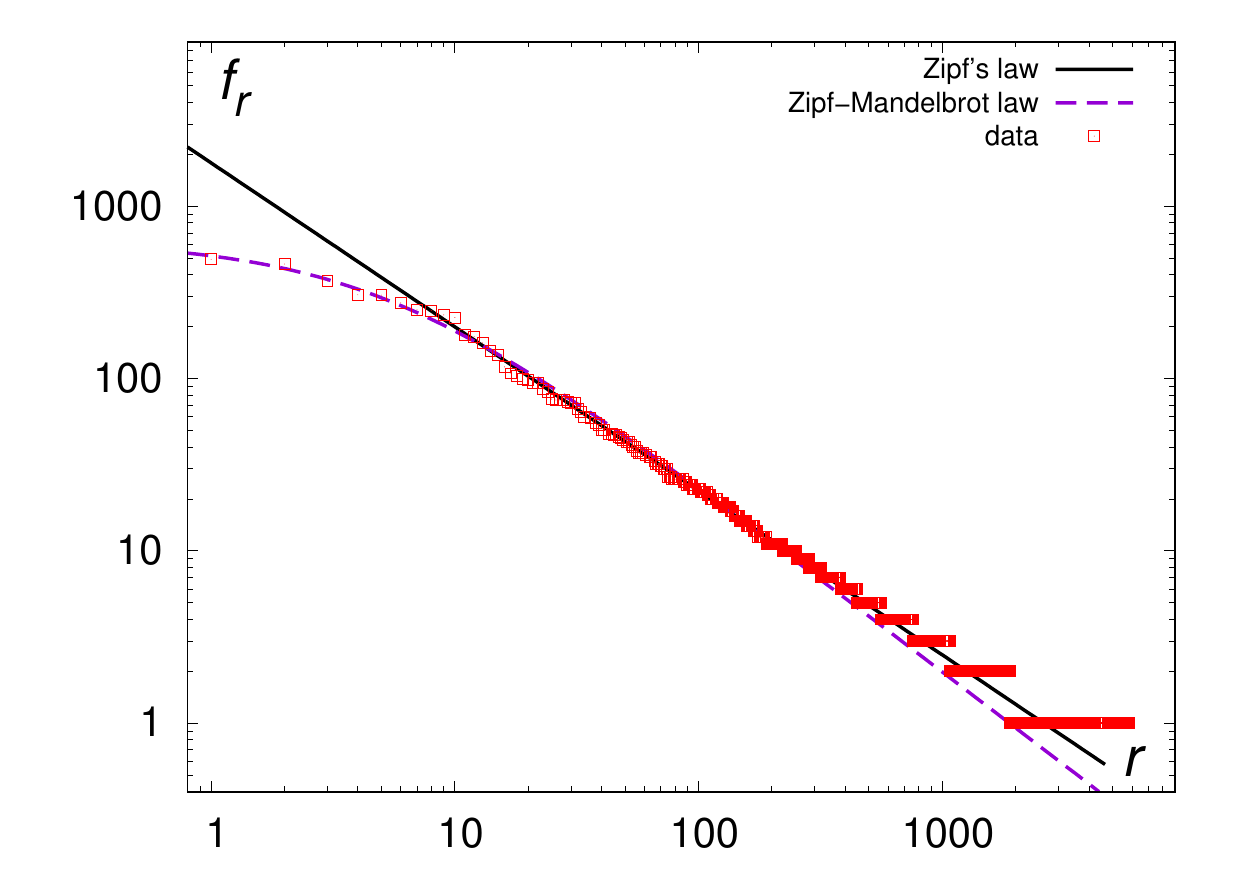}}
\vspace*{-1ex}
\caption{Rank--frequency dependence for the Ukrainian translation of \textit{Alice's Adventures in Wonderland} fitted using Zipf's law (\ref{eq:Zipf}) and the Zipf--Mandelbrot law (\ref{eq:ZM}).
The fitting parameters for (\ref{eq:Zipf}) are $A = 1786\pm7$, $z = -0.952 \pm 0.001$, the fitting range is $r\ge10$. The fitting parameters for (\ref{eq:ZM}) are as follows:
$C = 3524 \pm 50$, $M = 4.91 \pm 0.05$, $B = 1.083 \pm 0.004$ (own results)}
\label{fig:AW-ukr-Zipf}
\end{figure}

In Fig.~\ref{fig:AW-ukr-Zipf}, fitting results with models (\ref{eq:Zipf}) and (\ref{eq:ZM}) are compared. Here and further in this Chapter, fitting is done using the nonlinear least-squares Marquardt--Levenberg algorithm implemented in the \texttt{fit} procedure of \texttt{GnuPlot}, version 5.2.

Studies of large text collections (called \textit{corpora} being plural of \index{corpus (of texts)}\textit{corpus}) demonstrate a significant change in the value of the Zipfian exponent (from $z\simeq1$ to $z\gtrsim2$) at some high ranks that are not typically achieved for smaller samples. Such a change corresponds thus to two regimes in rank--frequency dependences  that can be identified with different types of vocabulary \cite{Ferrer_i_Cancho&Sole:2001,Montemurro:2001}. The first one with $z\simeq1$ might correspond to some core vocabulary. In fact, such two regimes can be also observed even in smaller samples\cite{Buk&Rovenchak:2004,Rovenchak:2015d} but they are less pronounced. In Fig.~\ref{fig:AW-RF} we can observe a similar behavior for the Hawaiian data. Marcelo Montemurro\cite{Montemurro:2004} suggested the following model with two exponents to recover such a two-regime behavior:
\begin{align}\label{eq:Montemurro}
f_r = \frac{D}{\lambda r^x+(1-\lambda)r^y}.
\end{align}

To conclude this section, it is worth mentioning another model for rank--frequency dependence, known as the continuous model of the \index{Yule distribution}\textit{Yule distribution}
\begin{align}\label{eq:Yule}
f_r = F\frac{k^r}{r^b}.
\end{align}
Usually, this model works well if the number of types is not large, a few tens or perhaps hundreds. In particular, relevant linguistic units are letters, phonemes, or syllables (in languages with strong phonotactic restrictions), cf. Refs.~\refcite{Martindale_etal:1996,Li&Miramontes:2011,Rovenchak_etal:2018}.

\section{``Temperature'' in linguistics}\label{sec:Temp}

Occasionally, the notion of `temperature' is used in a wide abstract sense, without reference to any quantitative parametrization, cf.\ Anna Wierzbicka's `emotional temperature' \cite[p.~395]{Wierzbicka:1992}. A rather simple notion of `verbal temperature' or its modification, `infinitive temperature' as the fraction of verbs (or infinitives, respectively) is also known \cite{Yanakiev:1977,Sokolova:2012}.

Beno{\^\i}t Mandelbrot was perhaps the first to introduce the concept of temperature in linguistic studies referring to physical considerations. In 1951, he coined the term \index{temperature!cybern\'etique}\textit{temp\'erature cybern\'etique du message} \cite{Mandelbrot:1951} (\index{temperature!informational}\textit{the informational temperature of the text} \cite{Mandelbrot:1953}) for the quantity $\theta=1/B$ in the rank--frequency dependence model (\ref{eq:ZM}).

Leaping several decades, we arrive at the beginning of the 21st century when several loosely related approaches involving temperature were suggested by various authors.

Kosmidis \textit{et al.}\cite{Kosmidis_etal:2006} introduce a measure of a word usefulness $k$. As the authors explain, it is related to the extent the word is essential for survival of an organized group of people. Postulating the respective (single-word) Hamiltonian as $H(k)=\eps\ln k$ they obtain the probability to find a word with usefulness $k$ as
\begin{align}
p(k) = \frac{1}{Z}e^{-H(k)/k_{\rm B}T} = \frac{1}{Z}k^{-\eps/k_{\rm B}T},
\end{align}
with $Z$ being the partition function defined in a standard way. The ``temperature'' $T$ is considered to be `a measure of the willingness (or ability) that the language speakers have in order to communicate' \cite{Kosmidis_etal:2006}. Curiously, from the assumption that the usefulness of a word is roughly proportional to its rank, the following estimation for the Zipfian exponent can be obtained:
\begin{align}
z\simeq\frac{\eps}{k_{\rm B}T}.
\end{align}

\index{temperature!of text}
The Boltzmann distribution was later applied by Miyazima and Yamamoto \cite{Miyazima&Yamamoto:2008} within an innovative approach for the measurement of temperature of texts. The authors suggested that the relative frequency of a word in the American National Corpus (ANC) $p_{\rm word}$ can be expressed using the word's energy $E_{\rm word}$ and a parameter $T$ called `temperature':
\begin{align}
p_{\rm word} = e^{-E_{\rm word}/k_{\rm B}T}.
\end{align}
The Boltzmann constant is then set to unity $k_{\rm B}=1$ and the temperature of the ANC is assumed to be $T=100$\,K (with kelvins serving units for both energy and temperature). In this way, values of $E_{\rm word}$ are uniquely defined for every word based on its frequency in ANC. To measure the temperature $t$ of a given text, one should fit the observed word frequencies using the function
\begin{align}
p_{\rm word}^{\rm obs.} = Ae^{-E_{\rm word}/t}.
\end{align}
The values of $t$ calculated for some English texts vary between 65 and 148~K \cite{Miyazima&Yamamoto:2008}. It should be noted that only the 100 most frequent words are considered in the calculations.

A similar approach was later used by Chang \textit{et al.}\cite{Chang_etal:2017}. In that work, the so-called information-based energy was also introduced, with the definition involving the Shannon entropy, as opposed to total energy defined in a standard way.

In 2010, we suggested another model involving the temperature parameter, which was based on the Bose--Einstein distribution \cite{Rovenchak&Buk:2011PhysA,Rovenchak&Buk:2011JPS}. The underlying idea was to fit the observed frequency spectrum using the function
\begin{align}\label{eq:Nj-exp}
N_j = \frac{1}{z^{-1}e^{\eps_j/T}-1},
\end{align}
where the ``fugacity'' $z$ is fixed by the number of \textit{hapax legomena}
\begin{align}\label{eq:zN1}
z = \frac{N_1}{N_1+1}
\end{align}
and the spectrum is chosen in a simple form
\begin{align}\label{eq:j-1}
\eps_j=(j-1)^\alpha, \qquad 1<\alpha<2.
\end{align}

In such a model, only several lowermost $j$s contribute significantly as $N_j$ rapidly drops at large values of $\eps_j$. For practical purposes it is thus reasonable to fit the observed data using (\ref{eq:Nj-exp})--(\ref{eq:j-1}) via two parameters, $\alpha$ and $T$, for $j=2\div j_{\rm max}$ [note that $j=1$ is used to fix $z$, see Eq.~(\ref{eq:zN1})].

To apply some ``natural'' definition for the upper limit $j_{\rm max}$ one can use the so-called $k$-point, which is the largest value of $j$ satisfying $N_j\ge j$. It is suggested that the $k$-point roughly marks a border in the vocabulary between high-frequency synsemantic (auxiliary or function) words and low-frequency autosemantic (meaningful) words \cite[p.~37]{Popescu_etal:2009}. Typical values of $j_{\rm max}$ in our studies are 10--15.

\begin{figure}
\centerline{\includegraphics[scale=0.8]{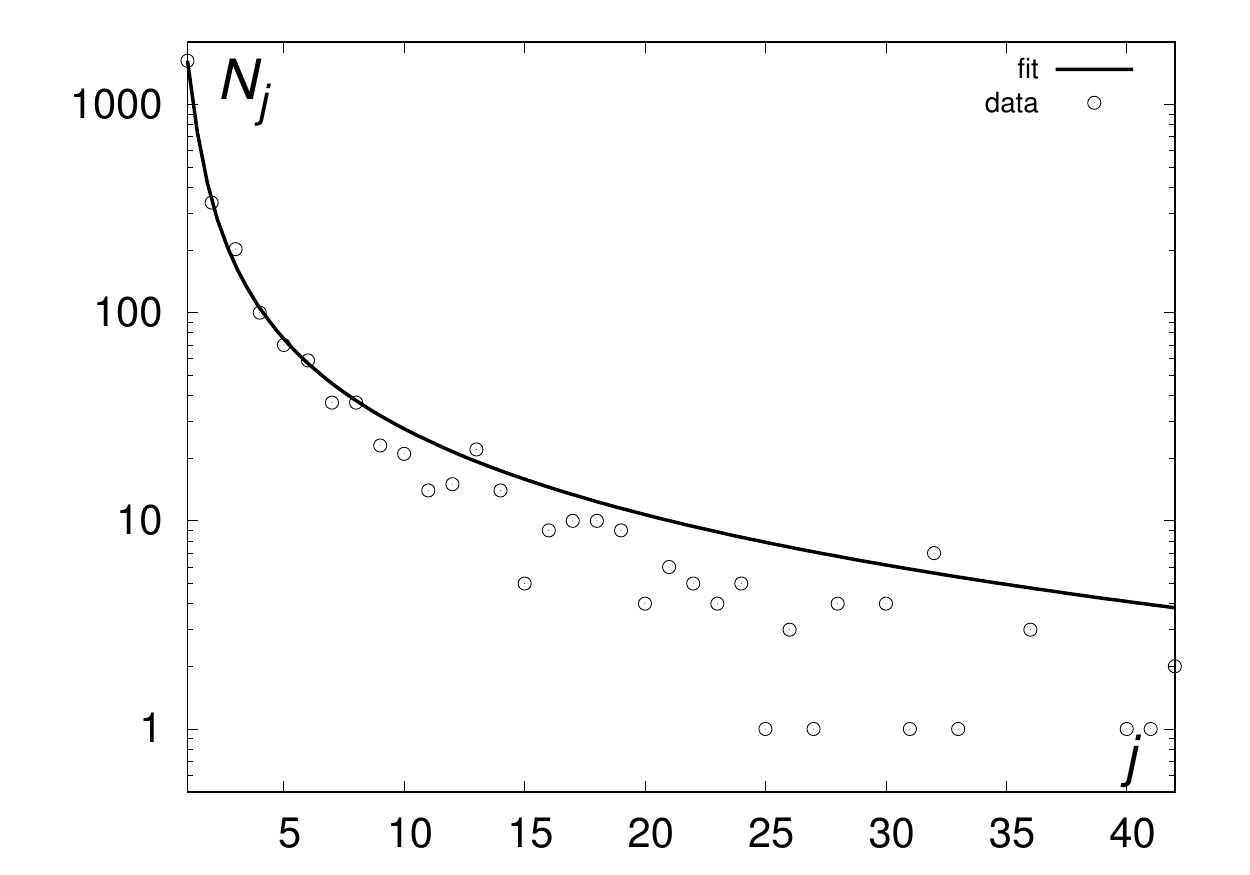}}
\vspace*{-1ex}
\caption{Frequency spectrum for the Swahili translation of \textit{Alice's Adventures in Wonderland} fitted using (\ref{eq:Nj-exp}). The upper limit for fitting is $j_{\rm max}=14$. The values of the fitting parameters are as follows:
$\alpha = 1.25 \pm 0.06$, $T = 446 \pm 20$ (own results)}
\label{fig:AW-swa-Nj}
\end{figure}

An example of the fitting is shown in Fig.~\ref{fig:AW-swa-Nj} for the frequency spectrum obtained from the translation of \textit{Alice's Adventures in Wonderland} in Swahili.%
\footnote{For the original studies\cite{Rovenchak:2015SLT,Rovenchak:2015CSCS} with this novel, various open sources were used to collect texts in different languages during 2013--2015. Those websites are mostly down nowadays, unfortunately. As of April 2022, the Swahili translation can be located at \url{https://www.maktaba.org/download/file/538}.}

The approaches from Ref.~\refcite{Miyazima&Yamamoto:2008} and Refs.~\refcite{Rovenchak&Buk:2011PhysA,Rovenchak&Buk:2011JPS} were recently applied to the analysis of \textit{Ragas}, melodic elements in Indian classical music \cite{Roy_etal:2021}. The possibility to use the obtained parameters in the categorization of Ragas was demonstrated.

As we will see in subsequent Sections, the ``temperatures'' from (\ref{eq:Nj-exp})--(\ref{eq:j-1}) can be related to certain properties of language grammar. Similar observations with other approaches described above concern, in particular, relations of the respective ``temperatures'' and vocabulary size growth in children with age\cite{Kosmidis_etal:2006}, changes in textbooks of English as a foreign language between junior and senior high schools in Japan\cite{Miyazima&Yamamoto:2008}, and even a measure of writer's creativity\cite{Chang_etal:2017}.

\section{``Temperature'' scaling}\label{sec:Tscaling}

Even superficial analysis shows that the parameter $T$ significantly depends on the number of tokens $N$. From the first glance this might be surprising as in physics temperature is an intensive property, i.\,e., not depending on the number of particles in the system. A way to drive the analogy back into the corral is, in fact, the thermodynamic limit notion.

Usually, for $N$ particles confined in a box of volume $V$, the thermodynamic limit is defined so that the ratio $N/V$ is kept constant while $N\to\infty$ and $V\to\infty$ simultaneously. No temperature enters this definition. But the situation becomes very different if particles are placed in an external potential. For the trapping potential in the form $U(r)\propto\omega r^n$, the thermodynamic limit condition reads \cite{Yukalov:2005} 
\begin{align}\label{eq:td_lim}
N\omega^s = {\rm const} \qquad\textrm{while}\quad  N\to\infty \ \ \textrm{and}\ \ 
\omega\to0,
\end{align}
where the exponent
\begin{align}
s = D\left(\frac{1}{2}+\frac{1}{n}\right)
\end{align}
is defined by the potential power $n$ and the space dimensionality $D$.

So, for instance, the critical (Bose condensation) temperature given by\cite{Yukalov:2005} $T_c\propto N^{1/s}$ depends indeed on the number of particles and is kept constant only if the thermodynamic limit conditions (\ref{eq:td_lim}) are held.

The model used to define the ``temperature'' parameter in Eqs.~(\ref{eq:Nj-exp})--(\ref{eq:j-1}) does not account for any thermodynamic limiting transition, hence it is quite natural to expect that $T$ grows as $N$ increases.

The scaling was empirically proved\cite{Rovenchak&Buk:2011PhysA,Rovenchak:2015SLT,Buk&Rovenchak:2016} to be described very precisely by a simple power-law dependence
\begin{align}\label{eq:T-scaling}
T = tN^\beta,
\end{align}
and the following parameter (weakly-dependent on $N$) was initially\cite{Rovenchak&Buk:2011PhysA,Rovenchak&Buk:2011JPS} used to make comparative analysis:
\begin{align}
\tau = \frac{\ln T}{\ln N}.
\end{align}

Empirically we have also discovered a strong inverse correlation between $\tau$ and $\beta$ \cite{Buk&Rovenchak:2016}. Assuming the power-law connection
\begin{align}
\beta = bt^{-a}\qquad\textrm{with}\quad a>0
\end{align}
we obtain immediately the link between $\tau$ and $\beta$ as follows
\begin{align}\label{eq:tau-beta}
\tau = \beta + \frac{1}{a}\frac{\ln b - \ln \beta}{\ln N}.
\end{align}
Obviously, in the limit of large $N$, the parameters $\tau$ and $\beta$ coincide. However, in practice the values of $N$ in the problems analyzed so far have been too small (typically, $N\sim10^4$) thus the logarithm being too weak to eliminate the contribution of the second term in \eref{eq:tau-beta}.

Here, it will be too hard to avoid deeping into physics for a moment (for the second time in this short Section!). Anyway, I will try to stay as shallowly as possible keeping in mind Altmann's warning mentioned in the Introduction. The physical analogy presented below was initially put forward in Ref.~\refcite{Rovenchak:2014}.

So, in thermodynamics the Gibbs free energy $G$ is related to the number of particles $N$ and the chemical potential $\mu=T\ln z$, where $z$ is the very fugacity parameter from \eref{eq:Nj-exp}, namely, $G=\mu N$. For the chemical potential one can easily obtain an expression via the number of hapaxes $N_1$:
\begin{align}
\mu=T\ln z = T\ln\frac{N_1}{N_1+1} \simeq -\frac{T}{N_1}
\end{align}
in the leading order assuming $N_1$ being large. From the nature of $G$ it follows that it is an extensive quantity, thus $G\propto N$. This means in turn that $\mu$ does not depend on $N$. If we then consider the scaling for the number of hapaxes\cite{Tuldava:1987} $N=cN^\gamma$ and the scaling of temperature (\ref{eq:T-scaling}) we immediately see that exponents $\beta$ and $\gamma$ should coincide. This relation really holds to a good accuracy \cite{Rovenchak:2014}.

\section{Nucleotide sequences}\label{sec:Genomes}

\begin{wrapfigure}{r}{0.36\textwidth}
\vspace*{-3ex}
\includegraphics[clip,height=0.26\textheight=1.0]{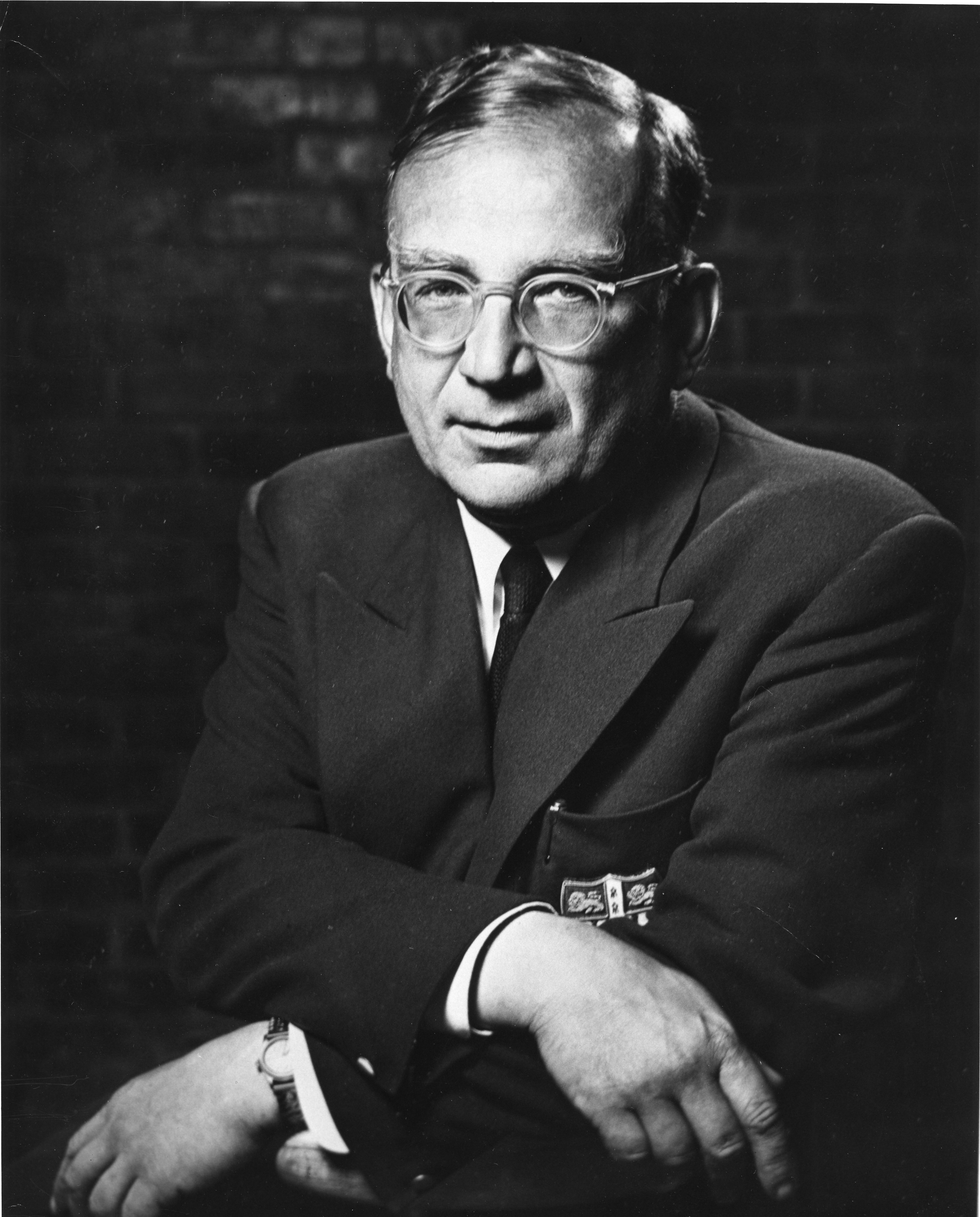}
\begin{centering}
{\footnotesize\qquad George Gamow\\[-0.2ex]
\hspace*{0.1em}
(04.03.1904 -- 19.08.1968)

}

\smallskip
{\scriptsize
\hspace*{0.1em}Courtesy: AIP Emilio Segr\`e\\
\hspace*{0.2em} Visual Archives, Physics \\
\hspace*{0.2em}Today Collection

}
\end{centering}
\end{wrapfigure}
While usual texts of words certainly exemplify complex systems of repeated tokens that are easy to collect and analyze, there are plenty of others. One can think, in particular, of genomes collected in large freely accessible databases\cite{NCBIH:www,Ensembl:www,HPA:www}. George Gamow\cite{Gamow:1954,Gamow:1954b,Gamow:1955} was likely the first\cite{Nanjundiah:2004} to suggest a linguistics analogy between nucleotide sequences in the molecules of deoxyribonucleic acid (DNA) and words in human languages, soon upon discovery of the DNA structure in 1953 by Watson \& Crick\cite{Watson&Crick:1953}. Note that a decade earlier, Erwin Schr\"odinger used the term ``hereditary code-script''\cite[Chap.~2]{Schrodinger:1944} with similar allusions\cite{Nanjundiah:2004}.
In this Section, I will describe some typical linguistic analogies used in DNA studies complementing them by my own recently suggested approach\cite{Rovenchak:2018} mostly applicable for comparatively short sequences.\index{DNA}

For integrity, it is worth recalling briefly the structure of DNA molecules. They are composed of two polynucleotide chains (\index{strand}\textit{strands}) made up of the following four nucleobases: adenine (A), cytosine (C), guanine (G), and thymine (T) \cite[p.~175]{Alberts_etal:2015}.
Nucleotides proper are formed as a nucleobase with a five-carbon sugar molecule (deoxyribose) and a phosphate group attached. Within a strand, strong covalent bonds occur between any bases while weaker hydrogen bonds between the strands have certain peculiarities. Only two types of such bonds are possible: adenine links with thymine and cytosine links with guanine. Such links form \index{base pair}\textit{base pairs} (bp) and their number is used to measure the length of DNA molecules.

\index{RNA}Ribonucleic acid (RNA) molecules have a slightly different structure comparing to DNA. The five-carbon sugar is ribose and there is a change in the set of four nucleobases: uracil
(U) in place of thymine (T) \cite[p.~39]{Alberts_etal:2015}. Unlike DNA, RNA molecules are mostly single-stranded. For the sake of homogeneity in data representation, in the NCBI databases the RNA structure uses T instead of U, and RNA sizes are quoted in bp (base pairs) instead of bases.

Genome sizes are often expressed in kilobases (1~kb = 1000~bp) or megabases (1~Mb = 1\,000\,000~bp). They vary in a very large range, from a few kilobases in case of viruses to over $10^5$~Mb in some plants \cite{Sessions:2013}. To achieve relative homogeneity of the analyzed data, further the study is limited to viral RNA and mammalian mitochondrial DNA\index{DNA!mitochondrial|see{mtDNA}} (mtDNA)\index{mtDNA}. The latter is a closed-circular, double-stranded molecule containing 16--17~kbp \cite{Taanman:1999}. 

In this Chapter, DNA and RNA building blocks are referred to as \index{nucleotide}\textit{nucleotides} and denoted by the respective nucleobase letter (A, C, G, or T) for simplicity. This does not cause any ambiguity or confusion for the purposes of the presented studies.

Nucleotide triplets, i.\,e., sequences of three nucleotides, are known as \index{codon}\textit{codons}. For instance, Leucine (Leu), which is an essential amino acid, is encoded by the following codons:
TTA, TTG, CTA, CTC, CTG, and CTT. Another essential amino acid, Valine (Val) is encoded by the codons GTA, CTC, GTG, and GTT. Certain combinations (TAA, TAG, and TGA) are stop codons \cite{Alberts_etal:2015}.

Larger structures include genes and polypeptides. Genes are typically composed of hundreds and thousands of base pairs but can achieve much larger sizes \cite{Wang_etal:2003,Schneider&Ebert:2004}. Polypeptides are composed of many amino acids, and one or more polypeptides further form proteins \cite{Kunugi:2014}.

Various approaches to establish the linguistics analogy in genome studies are attested.
In particular, Brendel \textit{et al.}\cite{Brendel_etal:1986} considered nucleotides as letters and $k$-mers  as words. Trifonov \textit{et al.} consider also frequently repeating nucleotide sequences (motifs) \cite{Trifonov_etal:2012}. Sungchul Ji\cite{Ji:1999,Ji:2020} lists analogy with more levels:
\begin{itemize}
\item letters -- 4 nucleotides or 20 amino acids;
\item words -- genes or polypeptides;
\item sentences -- sets of genes;
\item even ``punctuation'' as starting and terminating codons.
\end{itemize}

Focusing on relatively short genomes (mtDNA or virus RNA) one should look for some other units comparing to those mentioned above in order to achieve word-like behavior. Even for larger genomes, various deviations from Zipfian dependences in rank--frequency distributions of nucleotide $n$-tuples are observed in coding and non-coding DNA sequences \cite{Mantegna_etal:1995}.

The following approach was suggested \cite{Rovenchak:2018} to analyze relatively short genomes: substitute the most frequent nucleotide with a space and insert an empty sequence (denoted by X) between two consecutive spaces. For instance, given the human mtDNA (\textit{Homo sapiens}\cite{HomoSapiens:www})\label{page:ns-def}
\[
\textrm{GATCACAGGTCTATCACCCTATTAACCACTCACGGGA\ldots},
\]
where adenine (A) is the most frequent nucleotide, we obtain
\[
\textrm{G \ TC \ C \ GGTCT \ TC \ CCCT \ TT \ X \ CC \ CTC \ CGGG \ \ldots}\ .
\]
It appears that the resulting nucleotide sequences (``words'' between ``spaces'') exhibit significant similarities with words in natural languages in their rank--frequency distributions \cite{Rovenchak:2018,Husev&Rovenchak:2021a}. These very sequences are studied in Subsecs.~\ref{subsec:T-genomes} and \ref{subsec:S-genomes}.

\section{Some results with ``temperatures''}\label{sec:T-res}
In this Section, two rather different applications of approaches based on the ``temperature'' parameter are considered. The first problem deals with texts and the possibility to quantify some their properties relevant to the evolution of languages. In the second problem, mitochondrial genomes are considered within the linguistic analogy described above and fitting of respective frequency spectra with a modified version of distribution (\ref{eq:Nj-exp}) is studied.

\subsection{Language evolution}

Studies of several texts in a few dozen languages from very different language families\cite{Rovenchak&Buk:2011PhysA,Rovenchak&Buk:2011JPS,Rovenchak:2014,Rovenchak:2015SLT} revealed certain common properties of the parameters obtained while fitting frequency spectra within the procedure represented by Eqs.~(\ref{eq:Nj-exp})--(\ref{eq:j-1}). The main observation is that for languages with higher analyticity level (less word inflection) the values of $\tau=\ln T/\ln N$ and $\alpha$ are lower comparing to more synthetic languages. The reason of such a difference lies in different proportions of low-frequency words in case of analytic and synthetic languages---and indeed, these are low-frequency data that define the parameters in this ``temperature'' approach. Note that high-frequency vocabulary also behaves differently in analytic and synthetic languages, with the rank--frequency dependences being ``more Zipfian'' for the former, but this is not relevant for the analysis presented below.

Change of the parameters with time might be associated with a possibility to quantify the evolution of languages. Such an analysis requires texts produced at different language stages (for instance, Classical Greek and Modern Greek) or in historically related languages (Latin, Italian, Spanish). Naturally, significant differences can be expected for time lapses of at least several centuries. The best (and perhaps the only) option is to utilize religious texts. Fortunately, the availability of Anglo-Saxon version of the Gospel of John (first nine chapters)\cite{John:www} encouraged me to make studies of several language lineages\cite{Rovenchak:2014,Rovenchak:2019book}.

\begin{figure}[t]
\centerline{\includegraphics[scale=0.8]{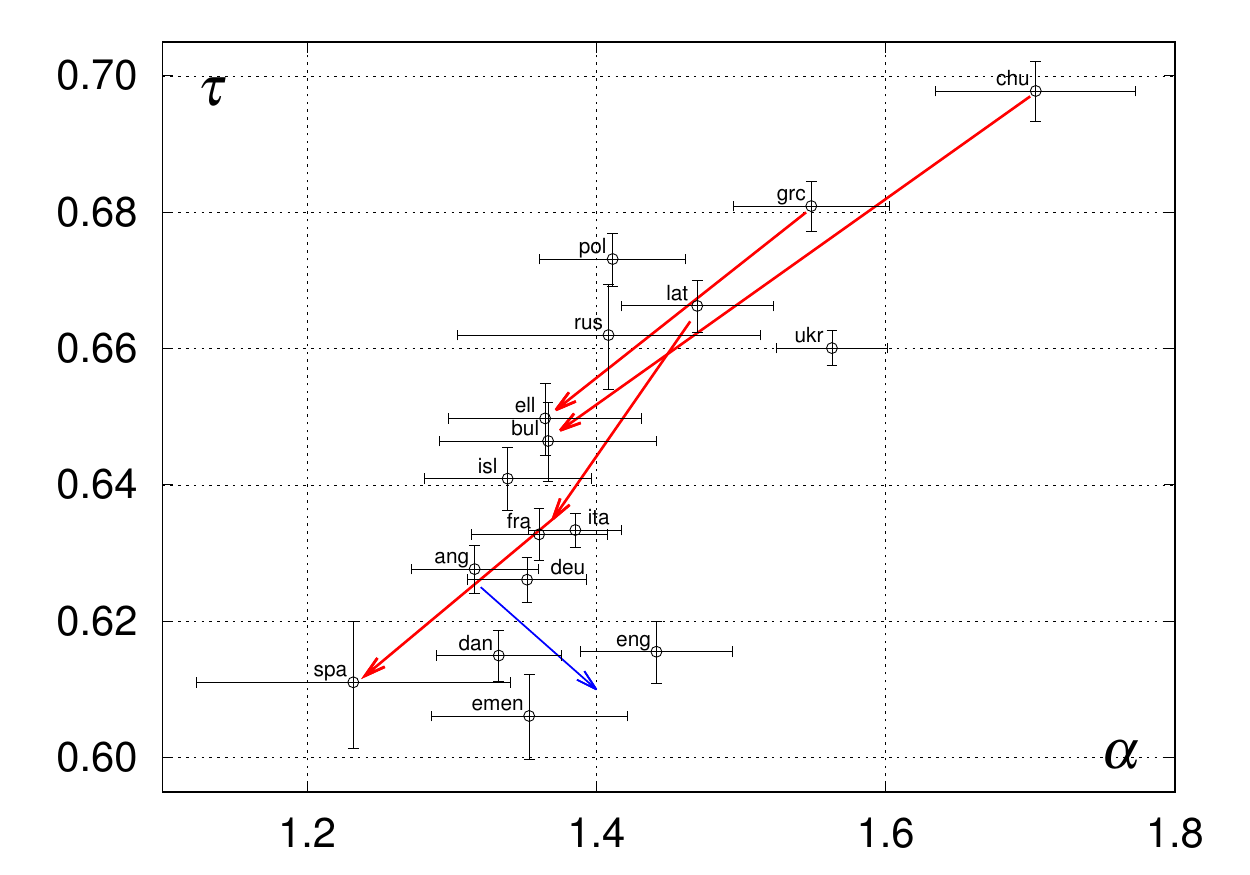}}
\vspace*{-1ex}
\caption{Translations of the Gospel of John in several languages (codes are given according to the ISO 639-3 standard\cite{ISO:www}, except for ISO 639-6 \texttt{emen} for the Early Modern English). Arrows demonstrate the evolution direction. Error bars show uncertainties in the determination of the fitting parameters. This figure contains partial data from previously published studies\cite{Rovenchak:2014,Rovenchak:2019book}}
\label{fig:NT9-T}
\end{figure}

The available texts%
\footnote{The texts were collected in 2012--2013 from open resources once available at  \url{http://unbound.biola.edu} and \url{http://gospelgo.com/bibles.htm} (both currently down). I~am able to provide raw plain text sources upon request.}
were processed ultimately yielding frequency spectra to be fitted according to the scheme given by Eqs.~(\ref{eq:Nj-exp})--(\ref{eq:j-1}). As the result, each version of the Gospel of John was characterized by two parameters, $\tau$ and $\alpha$. Figure~\ref{fig:NT9-T} shows results for several most interesting languages to be briefly discussed below.

Generally, the evolution is directed south west, form the domain of higher values of both $\tau$ and $\alpha$ to the domain of lower $\tau$ and $\alpha$. This is exemplified by the following lineages:
\begin{itemize}
\item Classical Greek (\texttt{grc}) to Modern Greek (\texttt{ell});
\item Latin (\texttt{lat}) to Spanish (\texttt{spa}) via French (\texttt{fra}) and Italian (\texttt{ita});
\item Old Church Slavonic (\texttt{chu}) to Bulgarian (\texttt{bul}) via the domain surrounded by Polish (\texttt{pol}), Russian (\texttt{rus}), and Ukrainian (\texttt{ukr}).
\end{itemize}

This supports the observations regarding the analytic and synthetic nature of language grammar described above. Moreover, we see that simplification of grammar with time is clearly seen. The best example is complex Old Church Slavonic and its close relative, modern Bulgarian, that almost lost noun inflection. Such results corroborate classical Greenberg's data\cite{Greenberg:1960}.
 
The opposite (south east) arrow joins the Anglo-Saxon (\texttt{ang}, Old English) language with the domain between the Early Modern English (\texttt{emen}) and Modern English (\texttt{eng}). For convenience, some other Germanic languages are shown: Icelandic (\texttt{isl}), German (\texttt{deu}), and Danish (\texttt{dan})---listed here roughly in the order of decreasing grammar complexity. This peculiarity in the English lineage might reflect a turbulent history of its interaction with other languages (including French). Still, the simplification of grammar is preserved at least in the decrease in the $\tau$ parameter.

\subsection{Distributions with deformed exponentials for genomes}\label{subsec:T-genomes}
The studies of frequency spectra obtained from rank--frequency distributions of nucleotide sequences (as defined at the end of \sref{sec:Genomes}, page~\pageref{page:ns-def}) in mammalian mtDNA\cite{Rovenchak:2018} and virus RNA\cite{Husev&Rovenchak:2021b} showed qualitative similarities with those of ordinary words. However, the decay of the genomic frequency spectra at large $j$ appeared not as fast as in texts. So, a modification of (\ref{eq:Nj-exp}) might be required to catch such a large-$j$ behavior more accurately. It was suggested\cite{Rovenchak:2018} to use the following model
\begin{align}\label{eq:Nj-X}
N_j = \frac{1}{z^{-1}X(\eps_j/T)-1},
\end{align}
where $X(t)$ is not the ordinary exponential but some other function.

The following two deformed exponentials were analyzed. The first one was the Tsallis $q$-exponential is defined as\cite{Tsallis:1994}
\begin{align}\label{eq:qexpTsallis-def}
e_q^x = \left\{
\begin{array}{ll}
\exp(x), &\textrm{for\ } q=1,\\[6pt]
[1+(1-q)x]^{1/(1-q)},\ &\textrm{for\ }q\neq1\ \textrm{and}\ 
1+(1-q)x > 0,\\[6pt]
0^{1/(1-q)}, &\textrm{for\ }q\neq1\ \textrm{and}\ 
1+(1-q)x \leq 0.
\end{array}
\right.\hspace*{-1em}
\end{align}
It is a well-known function in nonextensive formulations of statistical mechanics\cite{Abe&Okamoto:2001,Tsallis:2009} and is applied in modeling fat-tail distributions in various domains\cite{Douglas_etal:2006,Ochiai&Nacher:2009,Zhao_etal:2021}.

The second considered alternative is the kappa-exponential of Kaniadakis given by\cite{Kaniadakis:2001,Kaniadakis:2013}
\begin{align}\label{eq:kappa-exp-def}
\exp_\varkappa (x) = \left( \sqrt{1 + \varkappa^2 x^2} + \varkappa x \right)^{\frac{1}{\varkappa}}.
\end{align}
This function also appears in studies of complex systems to account for power-law distributions\cite{Wada&Scarfone:2015,Kaniadakis_etal:2020,daSilva_etal:2021}.
Both deformed exponentials reduce to the ordinary exponential in the limits of $q\to1$ and $\varkappa\to 0$, respectively.

\begin{figure}[t]
\centerline{\includegraphics[scale=0.8]{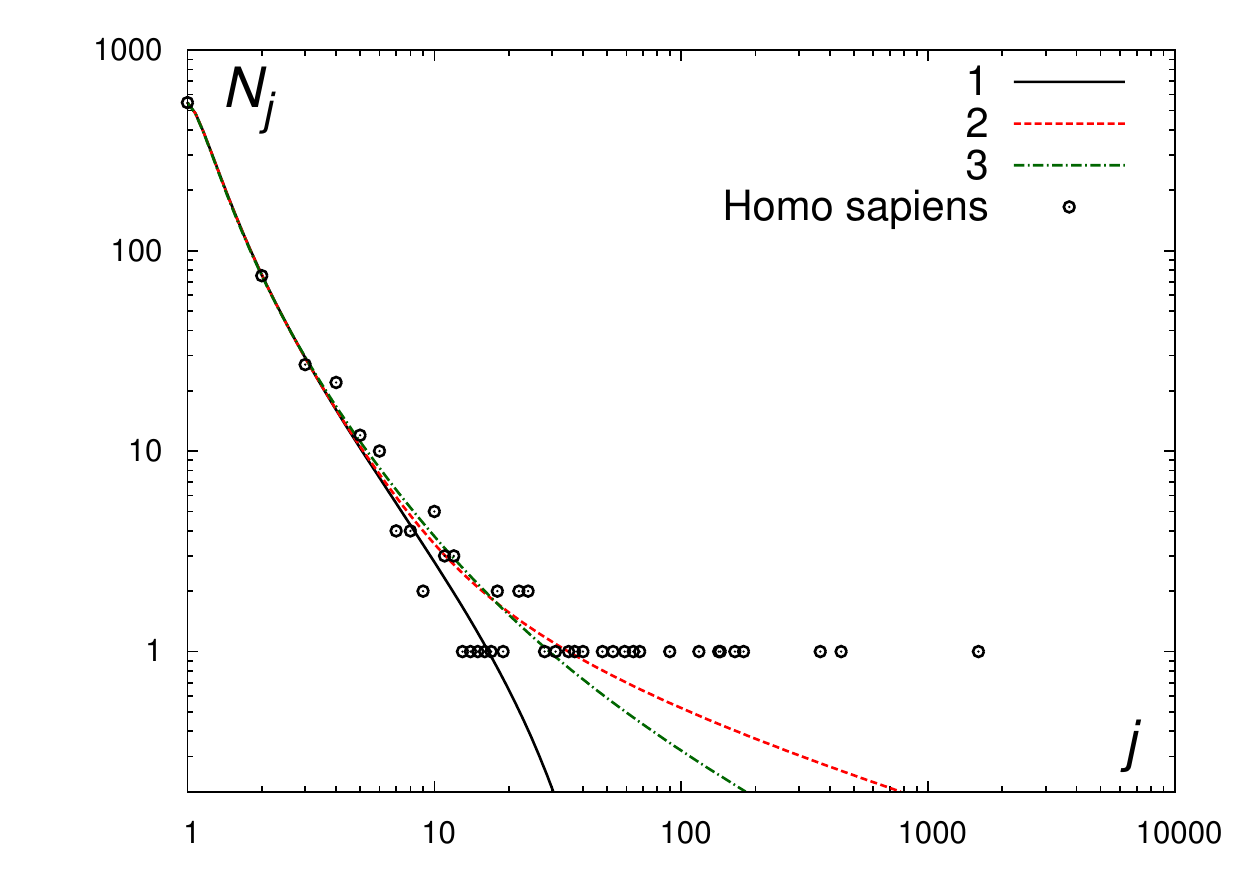}}
\vspace*{-2ex}
\caption{Fitting results for the frequency spectrum of nucleotide sequences in human mtDNA. Curve 1 corresponds to the ordinary exponential ($T = 88.8 \pm 3.2$), curve 2 corresponds to the kappa-exponential ($T = 88.4 \pm 1.6$, \ $\varkappa = 4.3 \pm 1.1$), and curve 3 stands for the Tsallis $q$-exponential ($T = 86.4 \pm 1.8$, \ $q = -1.4 \pm 0.8$); all the data are own results}
\label{fig:H-Homo_sapiens-spectrum}
\end{figure}

In \fref{fig:H-Homo_sapiens-spectrum}, results of fitting the frequency spectrum obtained for nucleotide sequences in human mitochondrial genome\cite{HomoSapiens:www} are shown. Three different curves correspond to $X(t)$ in \eref{eq:Nj-X} being the ordinary exponential, the Tsallis $q$-exponential (\ref{eq:qexpTsallis-def}), and the kappa-exponential (\ref{eq:kappa-exp-def}). All the calculations are made at the fixed value of $\alpha=1.5$ in $\eps_j = (j-1)^\alpha$. The value of $z$ is calculated using the number of hapaxes $N_1 = 548$. Negative values of $q$, while surprising at first glance, are not uncommon in studies involving the Tsallis $q$-exponential\cite{Costa_etal:1997,Nakamichi_etal:2002,Cerqueti_etal:2020}.

When similar calculations were made for some mammalian mtDNA\cite{Rovenchak:2018}, the mean value of $\varkappa$ for the \textsl{Felidae} family appeared to be larger than that of the \textsl{Ursidae} ($\langle\varkappa\rangle_{\rm F} = 6.3$ versus $\langle\varkappa\rangle_{\rm B} = 5.1$) yielding the conclusion that cats generally have longer tails than bears {\large\smiley}.

The presented examples show potential applicability of parameters associated with ``temperature'' in discrimination of various complex systems (text and biological species), although in many cases as a supplementary tool only owing to high uncertainties in calculations. In the following Sections, the use of another physically inspired quantity---entropy---together with some other parameters derivable from rank--frequency distributions is demonstrated.

\section{Entropy and some other frequency- and length-related parameters}\label{sec:Entropy+params}

The concept of entropy can be traced back to 1850 when it appeared under different names in works by William Rankine and Rudolf Clausius\cite{Truesdell:1980,Styer:2019} and later developed by the latter. The term itself originates in the work of Rudolf Clausius\cite{Clausius:1865}. For historical reference, the Clausius's passage containing the discussion about the naming is reproduced in Fig.~\ref{fig:Clausius1865-p390}.

\noindent
\begin{minipage}{0.66\textwidth}
\begin{figure}[H]
\centerline{\includegraphics[scale=0.76]{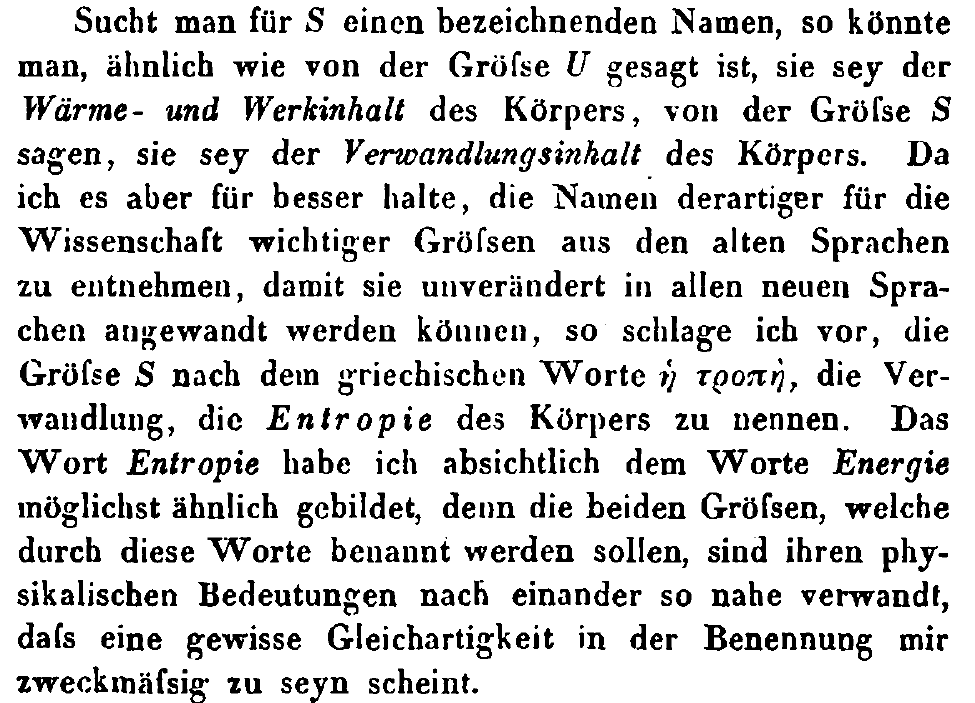}}
\vspace*{-1ex}
\caption
{The fragment of Clausius's 1865 paper\cite{Clausius:1865} (page 390)
 in \textit{Annalen der Physik und Chemie} 
containing the justification of the new term, \textit{Entropie}.}
\label{fig:Clausius1865-p390}
\end{figure}
\end{minipage}
\begin{minipage}{0.33\textwidth}
\vspace*{-3ex}
\includegraphics[clip,height=0.22\textheight=1.0]{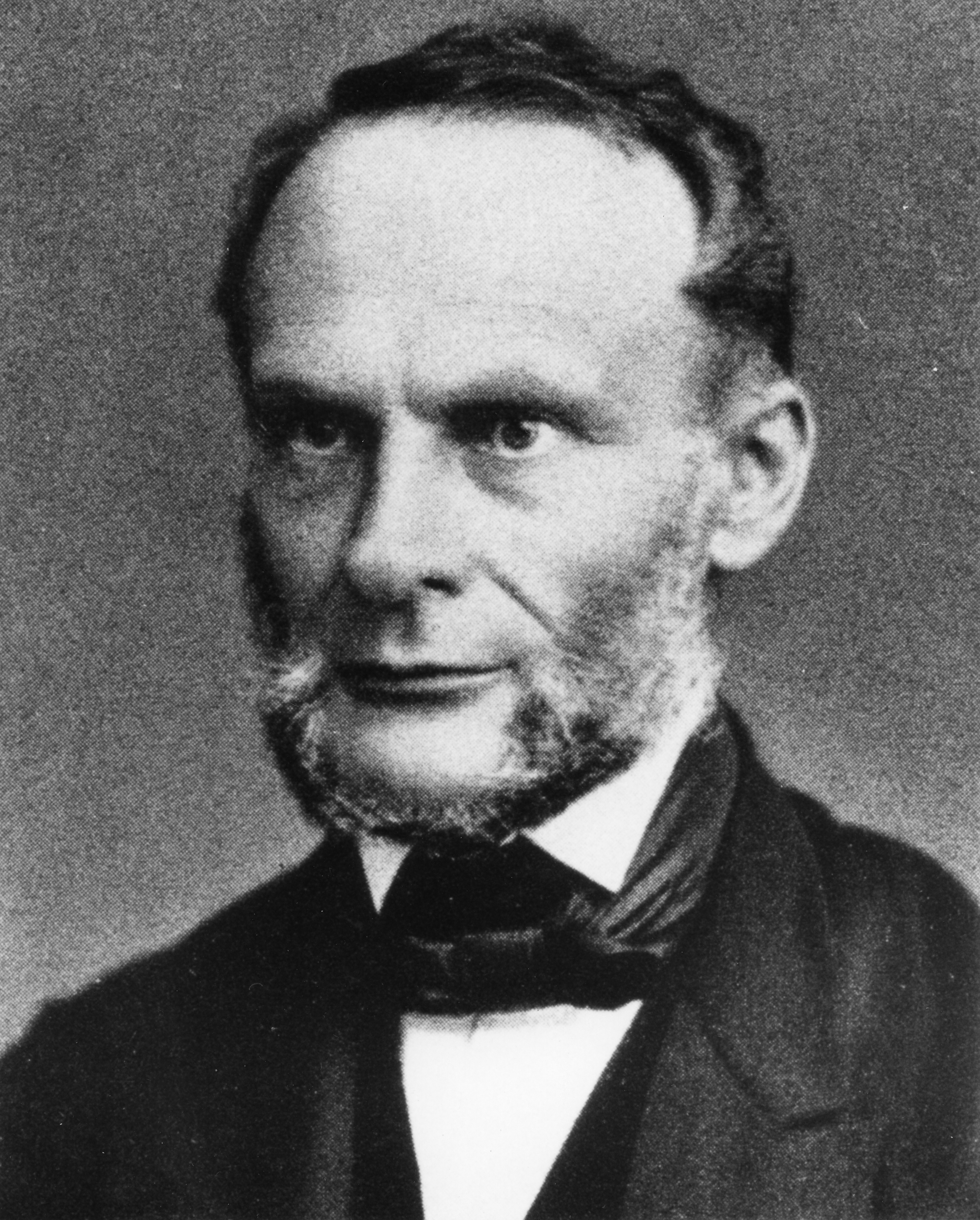}
\begin{centering}
{\footnotesize\qquad Rudolf Clausius\\[-0.2ex]
\hspace*{0.1em}
(02.01.1822 -- 24.08.1888)

}

\smallskip
{\scriptsize
\hspace*{0.1em}Courtesy: AIP Emilio Segr\`e\\
\hspace*{0.4em} Visual Archives,\\
\hspace*{0.8em}Lande Collection

}
\vspace*{-3ex}
\end{centering}
\end{minipage}

\bigskip\index{entropy}
A Google-assisted translation of the passage reads:
\begin{quote}
If one looks for a descriptive name for $S$, one could say, similarly to what is said of the quantity $U$, that it is the body's heat and work content, and of the quantity $S$ that it is the body's content of transformation. However, since I think it is better to take the names of such scientifically important quantities from the old languages so that they can be used unchanged in all new languages, I propose to use for the quantity $S$ the Greek word {\greektext <h trop'h}, the transformation, to name it the \textit{\,e\,n\,t\,r\,o\,p\,y\,} of the body. I deliberately formed the word \textit{entropy} as similar as possible to the word \textit{energy}, because the two quantities to be named by these words are so closely related in their physical meanings that a certain similarity in the naming seems to me to be expedient.
\end{quote}

\bigskip
Microscopic considerations toward the definition of entropy were started by Boltzmann\cite{Boltzmann:1877} in the 1870s but the well known relation $S=k\log W$ between entropy and the number of microstates $W$ was explicitly written for the first time by Planck\cite{Planck:1901} in 1901.

\begin{wrapfigure}{r}{0.36\textwidth}
\vspace*{-3ex}
\includegraphics[clip,height=0.22\textheight=1.0]{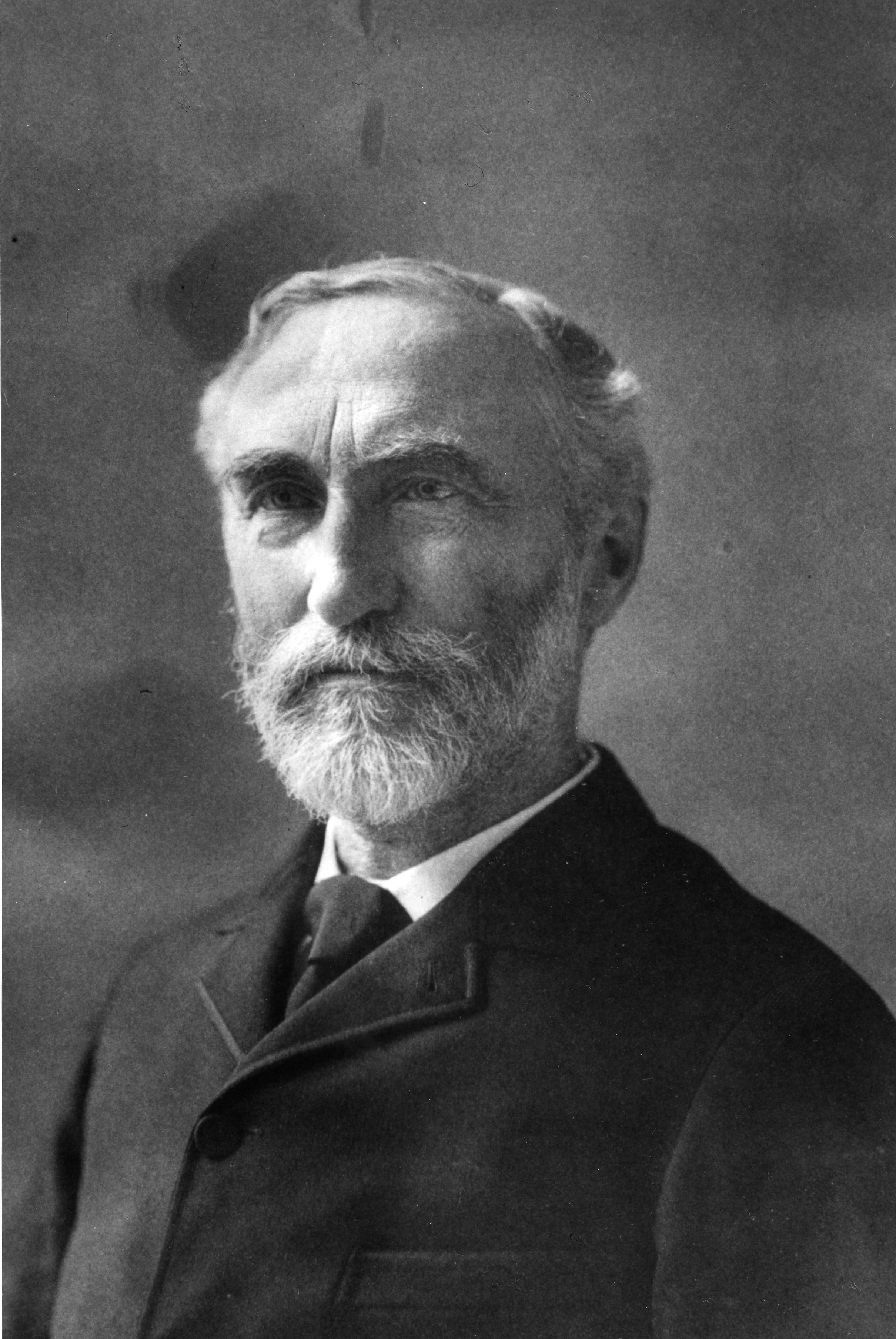}
\begin{centering}
{\footnotesize\qquad Josiah Willard Gibbs\\[-0.2ex]
\hspace*{0.1em}
(11.02.1839 -- 28.04.1903)

}

\smallskip
{\scriptsize
\hspace*{0.1em}Courtesy: AIP Emilio Segr\`e\\
\hspace*{0.4em} Visual Archives

}
\vspace*{-3ex}
\end{centering}
\end{wrapfigure}
\begin{figure}[h]
\centerline{\includegraphics[scale=0.8]{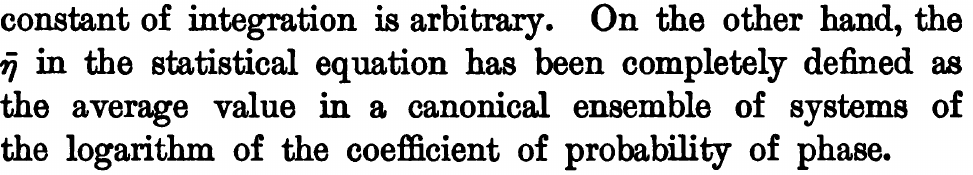}}
\caption{The fragment of Gibbs's 1902 book\cite{Gibbs:1902} (page 45) containing a passage related to the definition of entropy that is denoted by $\eta$. The book was digitized by Google and is available from \protect\url{https://archive.org}}
\label{fig:Gibbs1902-page45}
\end{figure}

Subsequent developments in the domain of statistical mechanics are also due to Josiah Willard Gibbs. It seems, however, that what we usually call the Gibbs formula for entropy, $S=-\sum p\ln p$ (or its equivalent expression for continuous variables via integral), had never been written by Gibbs himself. This expression can be obtained from the wording at the end of Chap.~IV of Gibbs' original book \cite{Gibbs:1902} published in 1902, see \fref{fig:Gibbs1902-page45}. The location of this expression appeared not an easy task though and was made thanks to Ref.~\refcite{Benguigui:2013}.

\bigskip

The concept of entropy was brought into information theory by Claude Shannon\cite{Shannon:1948a,Shannon:1948b}. For a discrete random variable $X$ achieving values $x_1,\ldots,x_M$ with probabilities $p_1,\ldots,p_M$, respectively, the following expression defines the entropy:
\begin{equation}\label{eq:Slog2-def}
H=-\sum_{r=1}^{M} p_r \log_b p_r ,
\end{equation}

\noindent
\begin{wrapfigure}{r}{0.36\textwidth}
\vspace*{-0ex}
\includegraphics[clip,height=0.22\textheight=1.0]{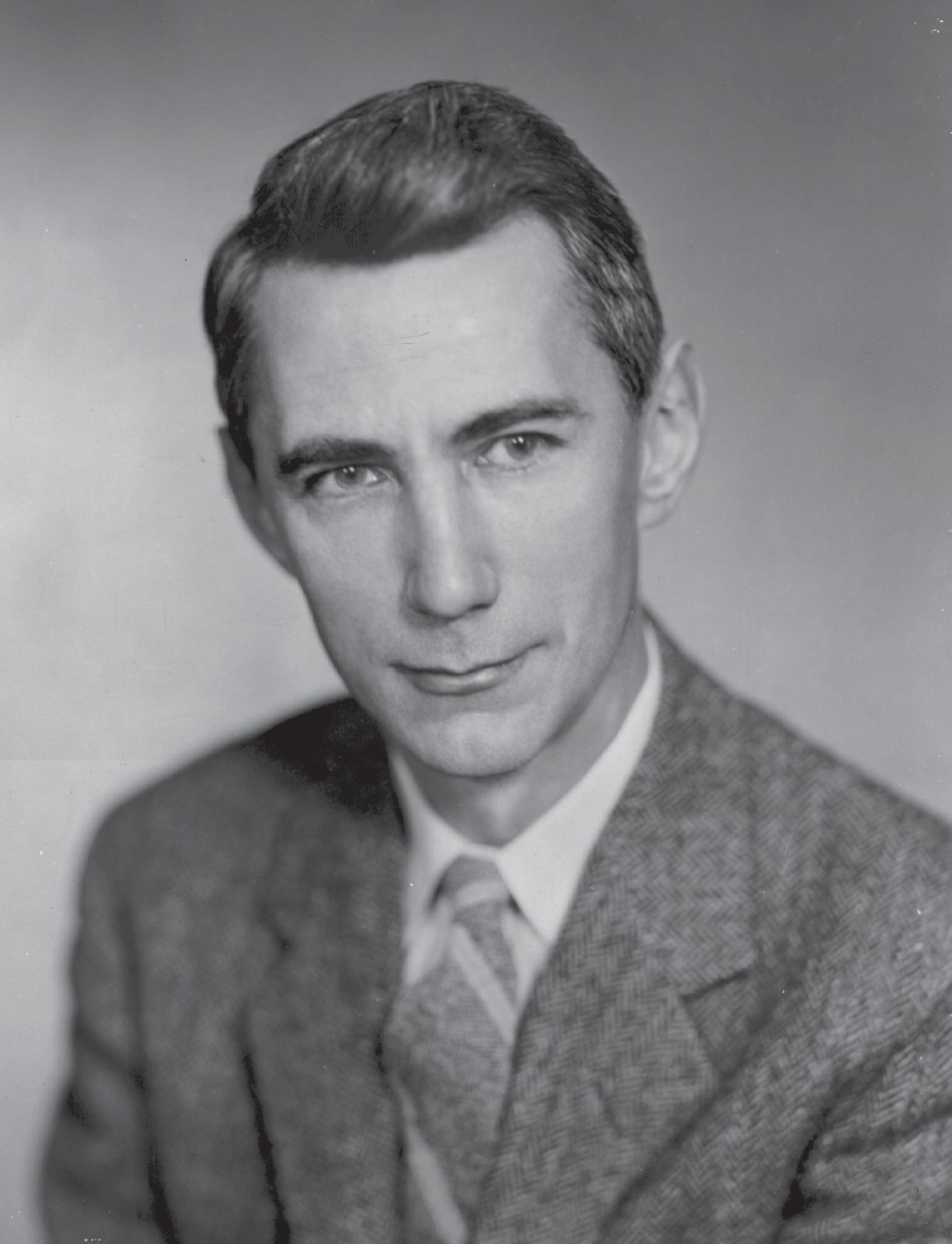}
\begin{centering}
{\footnotesize\qquad Claude Elwood Shannon\\[-0.2ex]
\hspace*{0.1em}
(30.04.1916 -- 24.02.2001)

}

\smallskip
{\scriptsize
\hspace*{0.1em}Source: Ref.~\refcite{James:2009}

}
\vspace*{-0ex}
\end{centering}
\end{wrapfigure}Aiming at measuring the information in \index{bit}\textit{bits}, one should use $b=2$; choosing the natural (base~$e$) logarithm one obtains information in \index{nat}\textit{nats} \cite{El-Haik&Yang:1999,You:2010}. Binary (base 2) logarithms are considered most common in information theory\cite{You:2010}. Shannon didn't specify explicitly the base in his original definitions but, for instance, from transformations given on page~630 of his original paper\cite{Shannon:1948b} one can conclude that he used the natural logarithm.

The use of entropy as a discriminating parameter might be rooted in its role as a measure of disorder\cite{Muller&Muller:2009,Styer:2019}. This was the main motivation for me to consider it---alongside other available quantities---in several studies devoted to the classification of complex systems\cite{Rovenchak:2018,Rovenchak&Rovenchak:2018,Husev&Rovenchak:2021a}.

Given a rank--frequency distribution, upon introducing the probability or relative frequency $p_r = f_r/N$, one can define entropy as follows: 
\begin{equation}\label{eq:S-def}
S=-\sum_{r=1}^{V} p_r \ln p_r ,
\end{equation}
where the upper summation limit corresponds to the total number of types.
Note that the normalization condition reads
\begin{equation}
\sum_{r=1}^{V} p_r = 1 = \frac{1}{N}\underbrace{\sum_{r=1}^{V} f_r}_{=N}.
\end{equation}

There are also plenty of other indicators that can be obtained from rank--frequency distributions. Recalling that the number of tokens is $N$ and the number of types is $V$, one can summarize the following list of ten parameters\cite{Rovenchak&Rovenchak:2018}:
\begin{itemlist}
\item type-token ratio TTR = $V/N$;

\item number of \textit{hapax legomena} \ $N_1$;

\item number of \textit{dis legomena} \ $N_2$;

\item fraction of \textit{hapax legomena} \ $N_1/N$;

\item fraction of \textit{dis legomena} \ $N_2/N$;

\item relation of \textit{hapax} and \textit{dis legomena} \ $N_1/N_2$;

\item repeat rate $R$ \cite{Zornig_etal:2016}:\index{repeat rate}
\begin{equation}
R = \sum_{r=1}^{V} p_r^2 = \frac{1}{N^2}\sum_{r=1}^{V} f_r^2
\end{equation} 

\item relative repeat rate $R_{\rm rel}$
\begin{equation}
R_{\rm rel} = \frac{1-R}{1-1/V}
\end{equation}
 
\item h-point \index{h-point}:
\begin{equation}
\hspace{-1em}h = \left\{
\begin{array}{cl}
r &\ \textrm{if there exists a solution of $r=f_r$}\\[6pt]
\displaystyle\frac{f_1r_2-f_2r_r}{r_2-r_1+f1-f_2}& \ 
\textrm{otherwise; note that $f_1>r_1$ and $f_2<r_2$}
\end{array}
\right.
\end{equation}

\item index $a$ being the scaling coefficient for the h-point (Popescu et al 2009):
\begin{equation}
a = \frac{N}{h^2}
\end{equation}
 
\end{itemlist}

\bigskip
Several parameters can be calculated if a unit to measure the length of a token is introduced. This will be specified in the respective sections. The length-related parameters are:
\begin{itemlist}
\item mean token length (first central moment)
\begin{equation}\label{eq:m1def}
m_1=\frac{1}{N} \sum_{i=1}^N x_i,
\end{equation}
where the summation runs over all the tokens of the analyzed sample and $x_i$ is the length of the $i$th token.

\item dispersion of token length (second central moment)
\begin{equation}\label{eq:m2def}
m_2=\frac{1}{N} \sum_{i=1}^N (x_i-m_1)^2.
\end{equation}
 
\item dispersion quotient \index{dispersion quotient}
\begin{equation}\label{eq:ddef}
d = \frac{m_2}{m_1-1}
\end{equation}
 
\item fraction of tokens with length $n$ denoted as $s_n$.
\end{itemlist}
 
It was shown that some of the parameters listed above are relevant in genre or author attribution of texts. These include, for instance, $d$ and $s_4$ considered in Ref.~\refcite{Kelih_etal:2005} or $R_{\rm rel}$ as one of the parameters for multivariate analysis\cite{Zornig_etal:2016}.
 
With the above defined quantities at hand, we will be able to experiment a bit in order to define their potential in discriminating complex systems. And what's more, in some cases nearly straightforward physics-based interpretation become possible as is demonstrated in the next Section.

\section{Some results with entropy et alii}\label{sec:S-res}
In order to direct the abstract discussion on entropy into a more settled down way, it is worth considering a toy example of a text with six tokens. Depending on the number of types we can calculate entropies as shown in \tref{tab:S}.

\begin{table}[h!!!]
\tbl{Entropies for different six-token texts}
{\begin{tabular}{@{}lll@{}} \toprule
Text & Types ($f_i$) & Entropy\\ 
\colrule
A B C D E F: & A (1) \ldots\ F (1) & 
	$S = -6\cdot\frac16\cdot\ln\frac16 = \ln 6 \simeq 1.79$\\[4pt]
A A A A A A: & A (6) & 
	$S = -1\cdot\ln 1 = 0$ \\[4pt]
A A A D E F: & A (3), D (1) \ldots\ F (1) & 
	$S = -\frac12\cdot\ln\frac12 - 3\cdot\frac16\cdot\ln\frac16 \simeq \mathbf{1.24}$\\[4pt]
A A A D D F: & A (3), D (2), F (1) & 
	$S = -\frac12\cdot\ln\frac12-\frac13\cdot\ln\frac13-2\cdot\frac16\cdot\ln\frac16 \simeq 1.31$\\ 
\botrule
\end{tabular}}
\label{tab:S}
\end{table}

\vspace*{-2ex}
So, the highest entropy is obtained for the text containing all the types with the same frequency. Such a text is hardly informative. The opposite case, with all the tokens being the same, obviously yields the zero entropy. Indeed, it contains a single piece of information. Two remaining examples require some effort to extract information from the text. However, we can speculate that such effort is lower in the third case (with $S\simeq1.24$, given in boldface) comparing to the fourth one. Indeed, the third text most likely tells about A while in the fourth one we can hesitate between A and D. Of course, real-world practices are much more complicated as most frequent words are just function words (like articles or conjunctions). But if I read a text in a foreign language knowing only a dozen words and I spot there six hyenas, one hare, and one lion, then most probably this is a tale with a hyena as the main character. The latter example demonstrates to some extent the role of entropy in extracting relevant information from a text.

In the following subsections, two problems are analyzed with respect to a possibility to discriminate two different kinds of complex systems---texts and viral RNAs---using entropy as one of the parameters.

\subsection{Measuring text comprehensibility}

Here, I would like to mention a situation that convinced me of the role of entropy in text comprehensibility. With some previous results for several genres in Ukrainian when the belles-lettres appeared to have the lowest entropy compared, in particular, to journalistic and scientific texts\cite{Buk&Rovenchak:2004}, it was even easier to come to such a conclusion. A friend of mine, who is an Austrian quantitative linguist, asked for some routine assistance with a text written in a language I am fluent in. Upon throwing a glance at one of the passages of the original and its translation into Ukrainian, my mother tongue, I had a strange feeling that the former was more ``readable''. This seemed really weird. So, I decided immediately to calculate the entropy of both versions---and indeed, it appeared \textit{higher} for the Ukrainian translation. Obviously, this fact has nothing to do with languages as a whole but is related to the translator's approach used.

Quantification of text comprehensibility is not an easy task as it involves a rather subjective estimation. This task becomes slightly simpler if a relative measure is sought for. Pursuing this aim, we have analyzed texts of a specific nature, namely, greetings from hierarchs of the Ukrainian Greek Catholic Church on the occasion of Christmas and Easter, the two most important Christian holidays\cite{Rovenchak&Rovenchak:2018}. Texts of such a kind are quite a typical subject of studies in quantitative linguistics \cite{Dai&Liu:2019,Kubat_etal:2021,Jiang_etal:2022}. 

Texts from three modern hierarchs were studied:%
\footnote{Texts of the greetings are accessible at the official websites \url{http://ugcc.ua} and \url{https://ugcc.lviv.ua} and eventually at other open resources.}
\begin{itemlist}
\item Lubomyr \textit{Cardinal} Husar M.S.U (Liubomyr Huzar, \ukrtext{Любомир Гузар}; 
*26.02.1933--\dag31.05.2017). He was Major Archbishop of Lviv in 2001--2005 and Major Archbishop of Kyiv and Halych in 2005--2011. The change in the title is due to the move of the see from Lviv to Kyiv.
\item Sviatoslav Shevchuk (\ukrtext{Святослав Шевчук};
*05.05.1970). His is the incumbent Supreme Archbishop of Kyiv and Halych, Head of the Ukrainian Greek Catholic Church since 2011.
\item Ihor Vozniak, C.SS.R. (Ihor Voznyak, \ukrtext{Ігор Возьняк}; 
*03.08.1952). He was Archbishop of the Lviv Archeparchy in 2005--2011, the Administrator of the Ukrainian Greek Catholic Church from February 10 to March 27, 2011, and is the incumbent Metropolitan of Lviv since 2011.
\end{itemlist}

Messages from Cardinal Husar are generally easier to perceive, and there might be various reasons for this: vocabulary, style, syntax, etc. The trial and error method allowed to pick several parameters from the list given in \sref{sec:Entropy+params} that might serve quantifiers of text comprehensibility\cite{Rovenchak&Rovenchak:2018}.

Length-related parameters ($m_1$, $m_2$, $d$, and $s_n$) were calculated using Eqs.~(\ref{eq:m1def})--(\ref{eq:ddef}) with $x_i$ being the word length in syllables, which is defined in Ukrainian very simply, as the number of letters for vowels 
$\langle$\ukrtext{а, е, и, і, о, у, є, ї, ю, я}$\rangle$ 
\cite{Macutek&Rovenchak:2011}. Note that some function words can consist of consonants only (\slantbox[0.24]{\ukrtext{в}} [v]
`in',
\slantbox[0.24]{\ukrtext{з}} [z]
`with, from', etc.) thus producing tokens of zero length.

The previously reported results\cite{Rovenchak&Rovenchak:2018} are supplemented here by Shevchuk's and Vozniak's greetings from 2018--2022. The original set of 36 texts was extended by 16 thus reaching 52 texts in the analysis presented here. The additions did not change the conclusions drawn in the preceding study. 

\begin{figure}
\centerline{\includegraphics[scale=0.8]{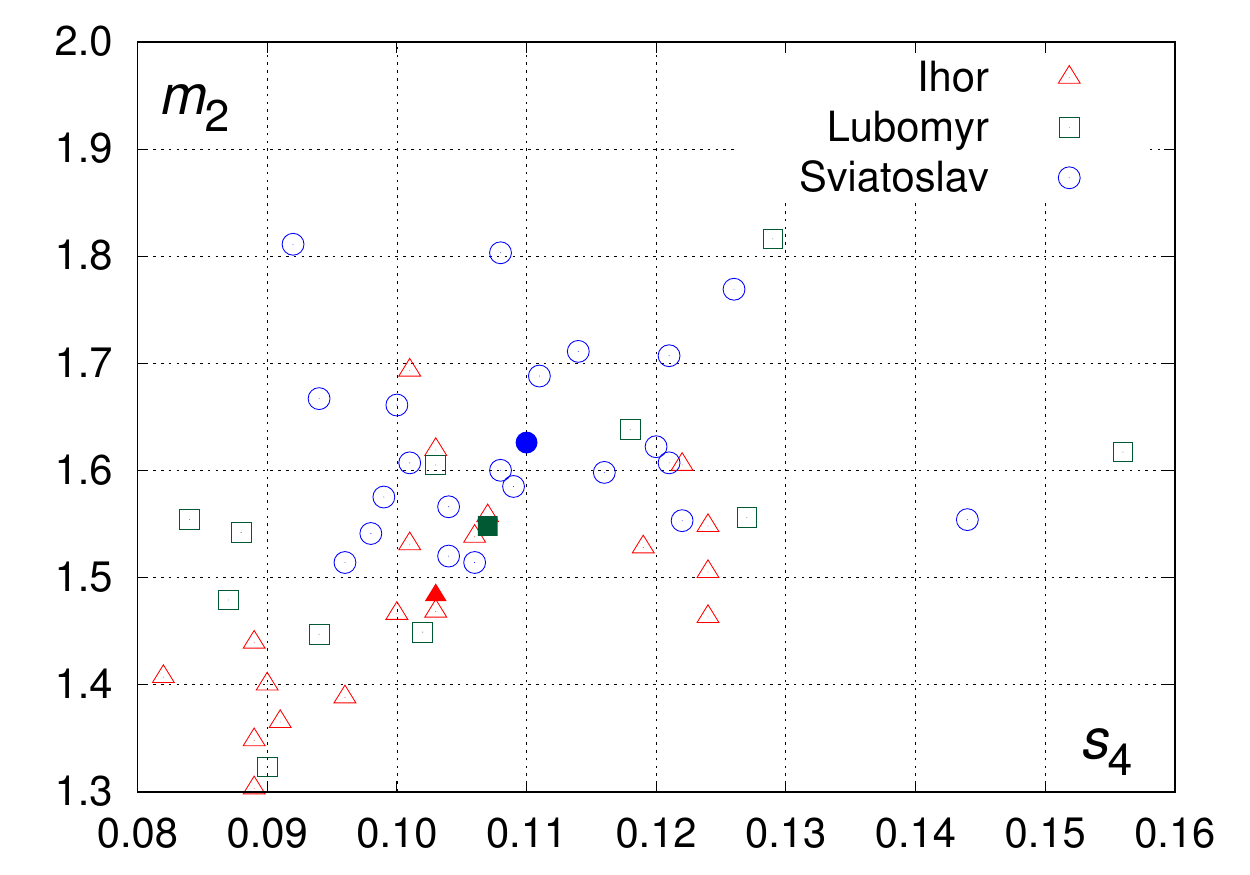}}
\vspace*{-2ex}
\caption{Text of greetings parametrized by the fraction of four-syllabic words $p_4$ and the dispersion of word lengths in syllables $m_2$. Mean values for the respective authors are marked with solid symbols. These are own results extending previously published work\cite{Rovenchak&Rovenchak:2018}}
\label{fig:addresses_UGCC-p4-m2}
\end{figure}

A bit unexpected result is that length-related parameters do not seem to have any significant role in quantifying text comprehensibility, cf.~\fref{fig:addresses_UGCC-p4-m2}. This contradicts, for instance, a na\"\i{}ve expectation that shorter words are easier to perceive.

\vspace*{-2ex}

\begin{figure}[h!!]
\centerline{\includegraphics[scale=0.8]{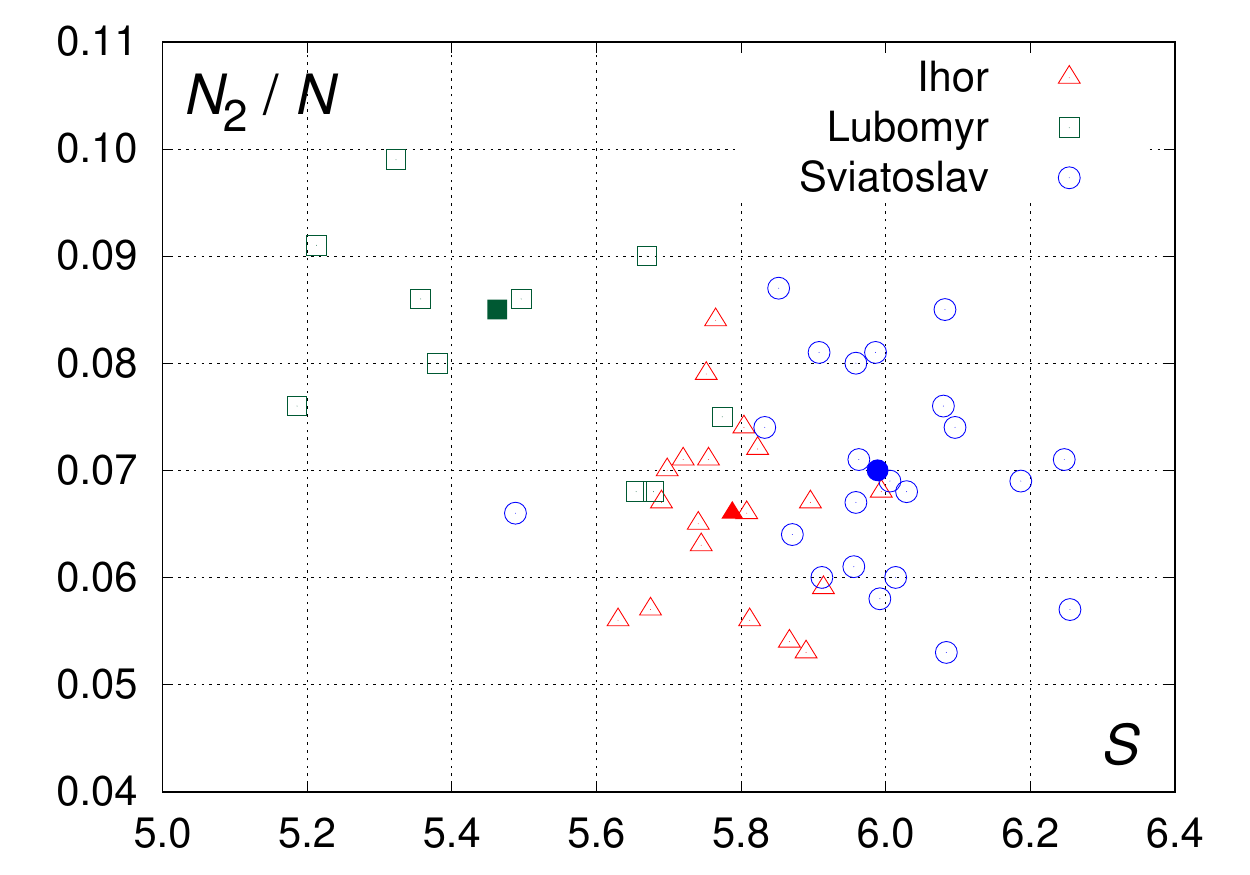}}
\vspace*{-1ex}
\caption{Text of greetings parametrized by entropy $S$ and the fraction of dis legomena $N_2/N$. Mean values for the respective authors are marked with solid symbols. These are own results extending previously published work\cite{Rovenchak&Rovenchak:2018}}
\label{fig:addresses_UGCC-S-N2N}
\end{figure}

\begin{figure}[h!!!]
\centerline{\includegraphics[scale=0.8]{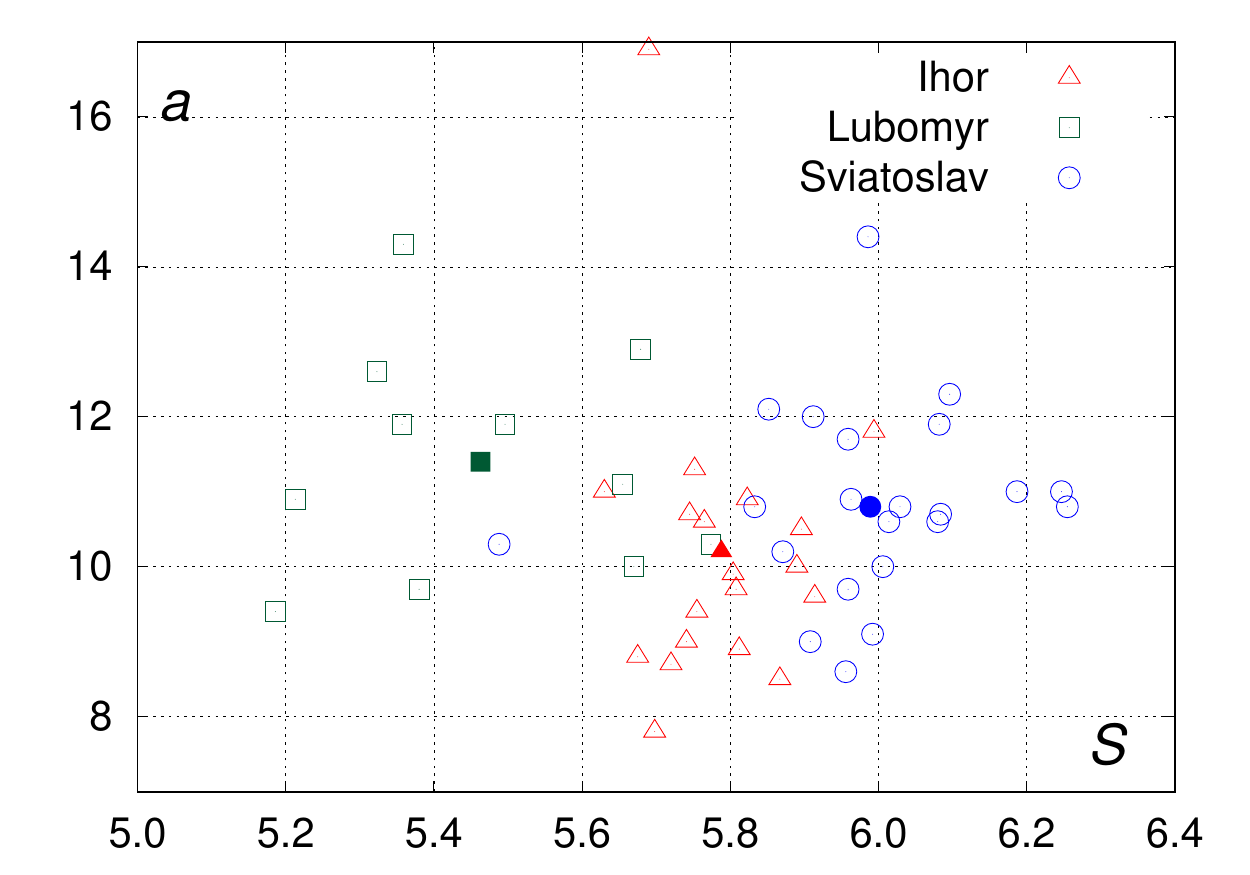}}
\vspace*{-1ex}
\caption{Text of greetings parametrized by entropy $S$ and index $a$. Mean values for the respective authors are marked with solid symbols. These are own results extending previously published work\cite{Rovenchak&Rovenchak:2018}}
\label{fig:addresses_UGCC-S-a}
\end{figure}

As one can see from Figs.~\ref{fig:addresses_UGCC-S-N2N}--\ref{fig:addresses_UGCC-S-a}, the studies texts demonstrate good separability with respect to entropy $S$, the fraction of dis legomena $N_2/N$, and the h-point scaling coefficient $a$. Note that the latter parameter was used for genre attribution by Z\"ornig \textit{et al.}\cite{Zornig_etal:2016}. The fraction of dis legomena in an attribution role was our result\cite{Rovenchak&Rovenchak:2018}.

The role of entropy as a discriminating parameter with respect to comprehensibility of texts is easily observed from the reported data: the lower the entropy the better comprehensibility. To a certain extent, one can think of a physical interpretation of such a result considering, for instance, a less ``disordered'' text to be easier to perceive. It is likely therefore that such ``(dis)ordering'' is reflected in the word frequency structure.

It should be stressed that weak or not separability of data with respect to some parameters just means they are not related to comprehensibility. Such parameters might still be useful for other purposes, for instance, in author attribution.

To complete this subsection, it might be interesting to establish a connection with one of the parameters related to comprehensibility and the ``temperature'' parameter discussed in Secs.~\ref{sec:Temp}--\ref{sec:Tscaling}. Indeed, the number of dis legomena can be expressed by \eref{eq:Nj-exp}:
\begin{equation}
N_2 = \frac{1}{z^{-1}e^{1/T}-1}.
\end{equation}
Considering the values of $T\gg1$ and $z\simeq1$, we easily obtain in the leading order 
\begin{equation}
N_2 \simeq T.
\end{equation}
Note that the ``temperature'' parameter is dimensionless, so the above formula makes sense. This is an interesting and initially unexpected connection between two major approaches to the classification of complex systems considered in this Chapter.


\subsection{RNA viruses, with focus on human coronaviruses}\label{subsec:S-genomes}
Now let us switch to a completely different subject of study, namely, to RNA viruses. This is done for the second time in this Chapter: previously, the ``temperature'' approach was discussed for mtDNA genomes in \sref{sec:T-res}. The ``complete difference'' between genomes and texts is, however, quite superficial: the former, as mentioned above in \sref{sec:Genomes}, can be considered within approaches rooted in linguistics. A major distinction between the two subjects is a possibility to justify the usage of entropy in quantifying comprehensibility of texts but virtually no chance to find a similar reasoning for rather abstractly defined nucleotide sequences in genomes.

Nonetheless, the study of mtDNA genomes revealed\cite{Rovenchak:2018} a possibility to use entropy and some length-related parameters calculated for nucleotide sequences to distinguish certain mammalian family and genera, at least tentatively. Similar analysis was then done for viral RNAs\cite{Husev&Rovenchak:2021a}. Its partial results extended to cover some new material are reported below. 

For nucleotide sequences defined at the end of \sref{sec:Genomes} (page~\pageref{page:ns-def}), with the most frequent thymine (T) nucleotide serving as a space, it is possible to calculate a set parameters, of which I focus on entropy $S$ and dispersion $m_2$ of mean token length $m_1$. In this case, lengths are measured as the number of nucleotides in the sequence but the empty sequence X is ascribed the zero length. 

\begin{figure}[h!]
\centerline{\includegraphics[scale=0.8]{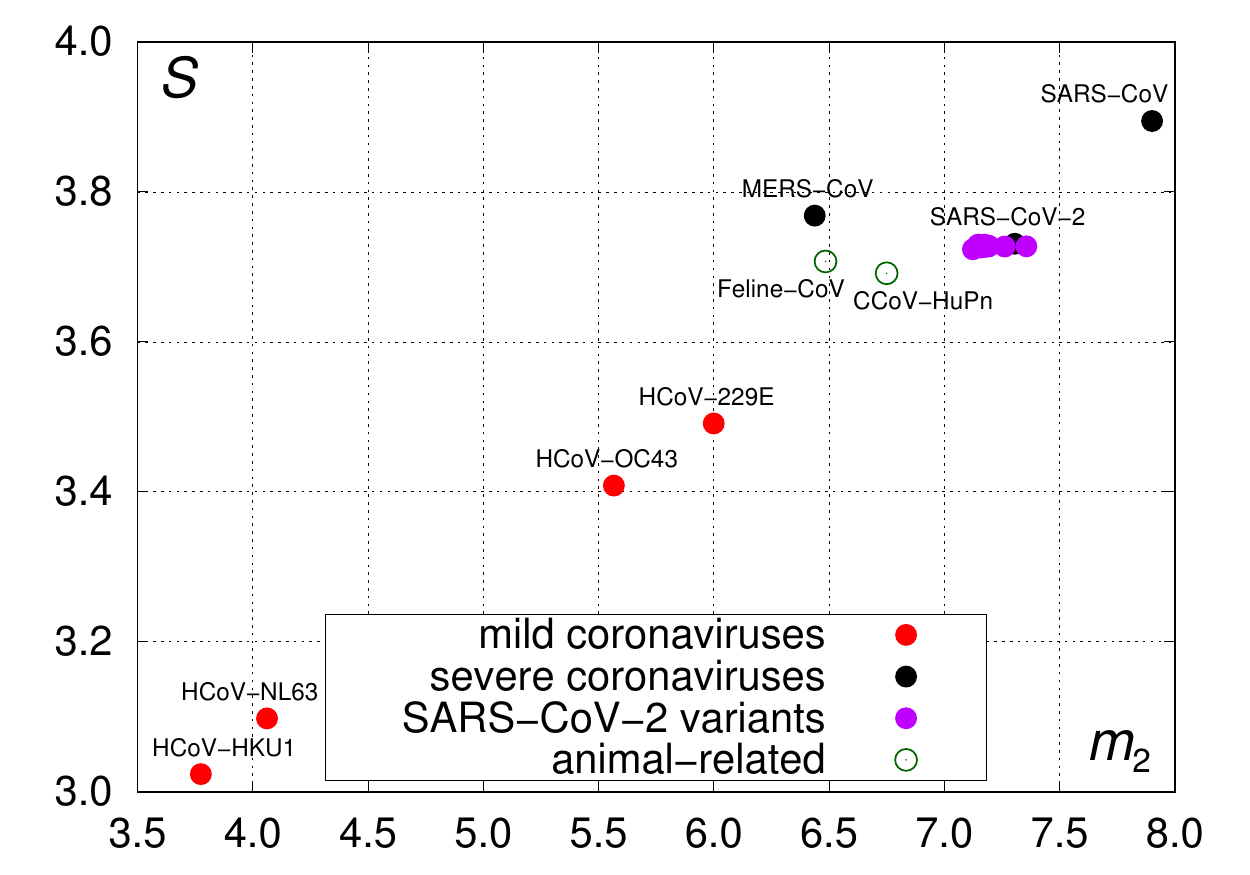}}
\vspace*{-1ex}
\caption{Coronaviruses on the $(m_2,S)$ plane. Solid circles correspond to human coronaviruses and open circles denote animal-related ones. Red filling marks viruses with associated mild diseases while black is for severe diseases. Violet color is used for SARS-CoV-2 variants. The previously published results\cite{Husev&Rovenchak:2021a} are extended by new data}
\label{fig:viruses}
\end{figure}

\tref{tab:viruses} lists the results of calculations for coronaviruses (mostly human\cite{Su_etal:2016,Wu_etal:2020}) being of particular interest in view of the recent COVID-19 pandemic\cite{Ciotti_etal:2020} due to severe acute respiratory syndrome coronavirus 2 (SARS-CoV-2). The respective data are plotted on the $(m_2;S)$ plane in \fref{fig:viruses}. Note that significant differences in the values of $m_1$ compared to the ratio between the genome size and the number of tokens $N$ is due to high abundance of zero-length X tokens.

\begin{table}[ht]
\renewcommand{\arraystretch}{1.06}
\tbl{Parameters for nucleotide sequences in human and some animal-related coronaviruses}
{\begin{tabular}{@{}lrrcccl@{}} \toprule
Virus	&	size, bp	&	$N$\hspace*{1em}	&	$S$	&	$m_1$	&	$m_2$	&	source$^{\text a}$	\\
\colrule
CCoV-HuPn	&	29086	&	9476	&	3.6914	&	2.0695	&	6.7500	&	MW591993.2	\\
Feline a-CoV	&	29355	&	9588	&	3.7072	&	2.0617	&	6.4845	&	315192962	\\
HCoV-229E	&	27317	&	9446	&	3.4910	&	1.8920	&	5.9998	&	12175745	\\
HCoV-HKU1	&	29926	&	12002	&	3.0233	&	1.4935	&	3.7750	&	85667876	\\
HCoV-NL63	&	27553	&	10806	&	3.0976	&	1.5499	&	4.0621	&	49169782	\\
HCoV-OC43	&	30741	&	10931	&	3.4081	&	1.8124	&	5.5669	&	1578871709	\\
MERS-CoV	&	30119	&	9800	&	3.7682	&	2.0735	&	6.4379	&	667489388	\\
SARS-CoV	&	29751	&	9144	&	3.8944	&	2.2537	&	7.9013	&	30271926	\\
SARS-CoV-2:	&		&		&		&		&		&		\\
\quad Reference	&	29903	&	9595	&	3.7307	&	2.1166	&	7.3059	&	NC\_045512	\\
\quad Alpha	&	29763	&	9569	&	3.7264	&	2.1105	&	7.1738	&	OL546784.1	\\
\quad Beta	&	29827	&	9588	&	3.7281	&	2.1110	&	7.1925	&	MZ314998.1	\\
\quad Gamma	&	29809	&	9582	&	3.7270	&	2.1110	&	7.1920	&	2056248244	\\
\quad Delta (AY)	&	29769	&	9586	&	3.7232	&	2.1056	&	7.1238	&	OM269121.1	\\
\quad Delta1 (B.1)	&	29836	&	9590	&	3.7269	&	2.1113	&	7.2616	&	OK091006.1	\\
\quad Eta	&	29700	&	9554	&	3.7256	&	2.1088	&	7.1562	&	MZ362439.1	\\
\quad Iota	&	29714	&	9540	&	3.7271	&	2.1148	&	7.3563	&	MZ702250.1	\\
\quad Omicron (BA)	&	29752	&	9561	&	3.7296	&	2.1119	&	7.1738	&	OM283600.1	\\
\quad Omicron-1	&	29788	&	9578	&	3.7295	&	2.1101	&	7.1465	&	OM095411.1	\\
\botrule
\end{tabular}}
\begin{tabnote}
$^{\text a}$Full source link is obtained by prefixing the listed sources with https://www.ncbi.nlm.nih.gov/nuccore/.
\end{tabnote}
\label{tab:viruses}
\vspace*{-1ex}
\end{table}

The worldwide circulation of numerous SARS-CoV-2 variants in 2020--2021 motivated the inclusion of its several strains and lineages in the analysis\cite{WHO:www,CDC:www}. The analyzed data also contain information about the canine coronavirus CCoV-HuPn-2018 recently discovered in a human patient \cite{Vlasova_etal:2022}, though with no human-to-human transmission reported. Feline coronavirus (Feline a-CoV, which is Feline infectious peritonitis virus) is given for future references.

The grouping of viruses on the $(m_2,S)$ plane in \fref{fig:viruses} is quite interesting. We can observe that higher entropies (and higher length dispersion values) correspond to species causing more severe diseases. The respective domain, however, includes also the two analyzed animal-related coronaviruses yielding thus limitations for direct usage of such severity distinctions based on values of $S$ and $m_2$. Fortunately, we would have here false positive alerts for the feline and canine coronaviruses, not false negative.

Previously,\cite{Husev&Rovenchak:2021a} we have also reported that severity of diseases caused by coronaviruses strongly correlates with the kind of the most frequent two-nucleotide sequence. Namely, it contains two identical nucleotides for coronaviruses linked to mild diseases and two different nucleotides for MERS and two SARS coronaviruses. The abovementioned CCoV-HuPn-2018 belongs to the first group, with GG being the most frequent two-nucleotide sequence (similarly to the feline coronavirus). So, let us hope that it will not be associated with any serious disease whatsoever.

The grouping of SARS-CoV-2 variants is very clearly seen in \fref{fig:viruses}. Such a result encourages the application of the suggested approach confirming in particular that the values of parameters are not just random, obtained upon deliberately chosen procedure of defining nucleotide sequences. One can expect such units to be another contribution in the linguistic analogies discussed in \sref{sec:Genomes}.

To conclude this Section, it might be interesting to mention that for nucleotide sequences discussed here rank--frequency distributions are essentially Zipfian, with wide plateaus at high ranks (low frequencies), thus indeed resembling words in natural languages. On the other hand, n-grams (to be specific, four-nucleotide chunks) in the studied genomes have qualitatively different rank--frequency distributions, closer to that of syllables\cite{Husev&Rovenchak:2021a}, and the Yule distribution (\ref{eq:Yule}) yields proper description for them instead of Zipf's law.

\section{Final remarks}\label{sec:Final}
The kaleidoscope of problems presented in this Chapter has several things in common. First of all, the considered approaches are inspired by physical models but transferred onto very different domains, far beyond physics. In this way, the used terminology often lacks initial physical sense and is mostly employed by analogy. Occasionally, however, reminiscences from physics arise nodding toward the desired unity of laws to describe Nature at very different levels, no matter how pathetic it may sound.

The second point is the kind of analyzed systems generally falling under the complex system category. Linguistics, additionally to physics, is a cross-cutting subject here as properties of texts and languages are accompanied by genomes---but considered within a linguistic analogy. The only tools required for the studies of texts are a pen and a computer, which is the luxury a physicist can rarely afford, as even theoreticians need experimental data to substantiate their research.

Another common issue in all the considered problems is that the attempted parametrization and discrimination of systems based on the parameter values yield only rather fuzzy domains. But there is a simple reason behind such results: we consider complex systems describable by a plethora of variables having very different origins. Whenever a two-parametric classification is attempted, a projection from this many-dimensional space is made onto a two-dimensional plane. To make further steps, other tools, like multivariate discriminant analysis, should be utilized.

The approaches described in this Chapter should nevertheless be treated as auxiliary tools in the classifications of complex systems. I do hope that certain hints about systems' properties and behavior might be obtained upon their applications, as was demonstrated by the example of language evolution, the quantification of text comprehensibility, and the analysis of RNA viral genomes. Further studies of other systems, the inclusion of new system types, and combinations of the developed techniques will decide.

\bibliographystyle{ws-rv-van}


\end{document}